\documentclass[sigconf]{acmart}

\newcommand{\revise}{\textcolor{black}}

\usepackage{graphicx}
\usepackage{array}

\usepackage{array}
\usepackage{amsmath}
\usepackage{footnote}
\usepackage{wrapfig}
\usepackage{bbding}
\usepackage{makecell}
\definecolor{DGcolor}{RGB}{0, 103, 156}
\definecolor{casecolor}{HTML}{D9D9D9}

\AtBeginDocument{%
  }

\setcopyright{acmlicensed}
\copyrightyear{2025} 
\acmYear{2025}
\setcopyright{cc}
\setcctype{by}
\acmConference[UIST '25]{The 38th Annual ACM Symposium on User Interface Software and Technology}{September 28-October 1, 2025}{Busan, Republic of Korea}
\acmBooktitle{The 38th Annual ACM Symposium on User Interface Software and Technology (UIST '25), September 28-October 1, 2025, Busan, Republic of Korea}
\acmDOI{10.1145/3746059.3747725}
\acmISBN{979-8-4007-2037-6/2025/09}

\acmSubmissionID{8158}

\begin{document}

\title{MEDebiaser: A Human-AI Feedback System for Mitigating Bias in Multi-label Medical Image Classification}


\author{Shaohan Shi}
\affiliation{%
  \institution{ShanghaiTech University}
  \city{Shanghai}
  \country{China}}
\email{shishh2023@shanghaitech.edu.cn}

\author{Yuheng Shao}
\affiliation{%
  \institution{ShanghaiTech University}
  \city{Shanghai}
  \country{China}}
\email{shaoyh2024@shanghaitech.edu.cn}

\author{Haoran Jiang}
\affiliation{%
  \institution{ShanghaiTech University}
  \city{Shanghai}
  \country{China}}
\email{jianghr2023@shanghaitech.edu.cn}

\author{Yunjie Yao}
\affiliation{%
  \institution{ShanghaiTech University}
  \city{Shanghai}
  \country{China}}
\email{yaoyj2024@shanghaitech.edu.cn}

\author{Zhijun Zhang}
\affiliation{%
  \institution{Shuguang Hospital Affiliated to Shanghai University of Chinese Traditional Medicine}
  \city{Shanghai}
  \country{China}}
\email{zjzhang2007@sina.com}

\author{Xu Ding}
\authornote{Quan Li and Xu Ding are the corresponding authors.}
\affiliation{%
  \institution{Shuguang Hospital Affiliated to Shanghai University of Chinese Traditional Medicine}
  \city{Shanghai}
  \country{China}}
\email{xu.ding2018@outlook.com}

\author{Quan Li}
\authornotemark[1]
\affiliation{%
  \institution{ShanghaiTech University}
  \city{Shanghai}
  \country{China}}
\email{liquan@shanghaitech.edu.cn}

\renewcommand{\shortauthors}{Shaohan Shi, Yuheng Shao, Haoran Jiang, Yunjie Yao, Zhijun Zhang, Xu Ding, and Quan Li}

\begin{abstract}
Medical images often contain multiple labels with imbalanced distributions and co-occurrence, leading to bias in multi-label medical image classification. Close collaboration between medical professionals and machine learning practitioners has significantly advanced medical image analysis. However, traditional collaboration modes struggle to facilitate effective feedback between physicians and AI models, as integrating medical expertise into the training process via engineers can be time-consuming and labor-intensive. To bridge this gap, we introduce \textit{MEDebiaser}, an interactive system enabling physicians to directly refine AI models using local explanations. By combining prediction with attention loss functions and employing a customized ranking strategy to alleviate scalability, \textit{MEDebiaser} allows physicians to mitigate biases without technical expertise, reducing reliance on engineers, and thus enhancing more direct human-AI feedback. Our mechanism and user studies demonstrate that it effectively reduces biases, improves usability, and enhances collaboration efficiency, providing a practical solution for integrating medical expertise into AI-driven healthcare.
\end{abstract}




\maketitle

\section{Introduction}
\par Artificial Intelligence (AI) has significantly enhanced healthcare providers' workflows in various areas such as electronic medical records management~\cite{Miotto2018,8086133}, clinical decision support~\cite{10.1007/978-3-031-48316-5_1,Sutton2020}, and particularly medical image analysis~\cite{gong2023diffusion,Bozyel2024,10.1145/3544548.3580682,10.1145/3313831.3376807}. AI systems aid in the analysis of medical images, including X-rays, CT scans, and MRIs, to detect lesions and abnormalities, thereby increasing diagnostic accuracy and efficiency~\cite{KHALIFA2024100146,8606641,MARTINNOGUEROL2021317}.

\par When developing AI systems for medical image analysis, machine learning (ML) practitioners (hereafter ``engineers'') rely on medical professionals (hereafter ``physicians'') to provide a substantial number of accurately labeled or even annotated images for training purposes. These images need to precisely represent various diseases or conditions, allowing AI models to effectively differentiate between different medical scenarios. In many medical contexts, a single image may exhibit features of multiple conditions, a phenomenon known as \textit{Multi-label Image}~\cite{KHAN202145,10.1145/2556288.2557011,6471714}. When building a system to handle multi-label medical images, engineers face two primary challenges inherent in medical datasets. The first challenge is \textit{Imbalanced Distribution}~\cite{10.1007/978-3-030-58548-8_10}. Common diseases are overrepresented in the data, while rare diseases have fewer cases, leading to an imbalanced distribution of labels. The second challenge is \textit{Label Co-occurrence}~\cite{8961143}. Certain symptoms often appear together, creating complex dependencies between labels. For instance, in chest X-rays, conditions like cardiomegaly, consolidation, and edema often occur simultaneously~\cite{9071248}, which can lead the model to confuse these co-occurring symptoms as variations of the same issue. These challenges introduce biases during training, making it difficult for the model to accurately differentiate between symptoms. This, in turn, affects the accuracy of diagnostic results and can undermine physicians' trust in AI systems.

\par
\begin{figure*}[h]
  \centering
  \includegraphics[width=0.8\linewidth]{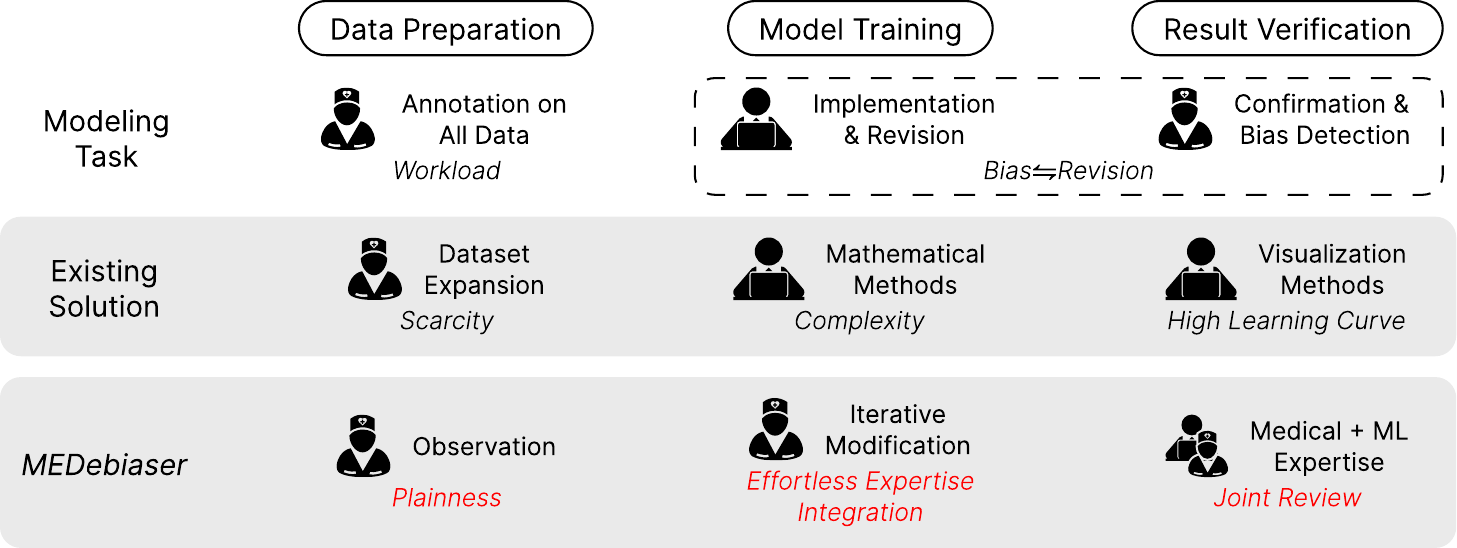}
  \caption{Traditional modeling tasks face challenges such as the high workload of annotating entire datasets and the inefficiencies of the ``Bias$\leftrightarrows$Revision'' mode. Existing solutions focus on dataset expansion---difficult to apply in the case of imbalanced data---or complex visualization and interaction systems. Additionally, integrating knowledge through engineers can be cumbersome and inefficient. \textit{MEDebiaser} overcomes these by providing an accessible visual interface that allows physicians to apply their expertise in familiar ways, ensuring continuous monitoring by both engineers and physicians and fostering seamless collaboration between the two.}
  \label{fig:traditional}
\end{figure*}

 \revise{The conventional approach to building MLMIC models involves a three-stage process: \textit{data preparation}, \textit{model training}, and \textit{result verification}~\cite{7930302} (\autoref{fig:traditional}). Physicians annotate images, engineers build and train models, and then physicians provide feedback on suboptimal results.} This ``Bias$\leftrightarrows$Revision'' cycle can be inefficient due to different work styles and perspectives, resulting in communication gaps and increased complexity in model iteration~\cite{MARTINNOGUEROL2021317}. While more advanced interactive and visualization tools have been developed to involve physicians more directly, they have their own drawbacks. Engineers often utilize visualization and interactive techniques to facilitate such feedback~\cite{A2023100230,yang2024foundation,ZHANG2024121282}, involving physicians in \textit{data preparation}~\cite{BORYS2023110786,LI2021345}, \textit{model training}~\cite{jcm8071050}, and \textit{result verification}~\cite{zhang2023ethics}. These approaches often include well-designed interactive features, but require physicians to grasp fundamental AI and ML concepts, creating a steep learning curve for those without a computer science background. Moreover, while physicians may find the model's explanations satisfactory, this does not necessarily ensure that the model remains practical as human knowledge continues to be integrated. Thus, joint evaluation by physicians and engineers is essential: physicians should review the model's interpretability for accuracy, while engineers analyze the data to ensure the model's ongoing usability.

\par \par The rapid integration of AI into medical image analysis has also fostered interdisciplinary collaboration between physicians and engineers~\cite{10.1007/978-3-031-48316-5_1,10.1145/3544548.3580945}. Physicians contribute valuable data and provide critical insights, while engineers leverage advanced AI technologies to apply this knowledge. This synergy not only amplifies the application of AI in the medical field but also promotes its widespread adoption, thereby revolutionizing the way medical diagnostics and treatments are approached~\cite{MARTINNOGUEROL2021317}. \revise{However, significant hurdles impede effective collaboration in MLMIC.} During \textit{data preparation}, engineers often rely on physicians to annotate various labels for each image in the dataset, i.e., specifying the features the ML system should learn to recognize. This approach, while essential, can be labor-intensive, especially when dealing with large datasets. In \textit{model training}, state-of-the-art models have shown remarkable performance~\cite{10350964,10350415,10350895,Park2023RobustAL,hong2023bagtrickslongtailedmultilabel,verma2023tamelongtailchestxray,10350884,10350374,10350833}, yet their direct application in clinical settings remains challenging due to physicians' limited ML expertise. Physicians typically engage passively, just relying on pre-developed solutions created by engineers~\cite{NASARIAN2024102412}. For \textit{result verification}, when model performance is suboptimal---such as struggles with symptom differentiation or misidentification---physicians provide feedback for engineers, helping to refine the model. Due to the scarcity of medical data, expanding the dataset is often difficult, leading engineers to use mathematical or algorithmic approaches to align physicians' expertise with model predictions. However, translating medical expertise into mathematical formulations is complex. 

\par \revise{To address these issues, our work focuses on two interconnected challenges in MLMIC: technical hurdles like imbalanced distributions and label co-occurrences, and the inefficient ``Bias$\leftrightarrows$Revision'' collaboration loops that result from workflow and perspective mismatches between physicians and engineers.} We introduce \textit{MEDebiaser} (\autoref{fig:traditional}), a human-in-the-loop system designed to bridge this gap by streamlining the revision cycle. This approach draws on the principles of mixed-initiative user interface design~\cite{10.1145/302979.303030}, aiming to achieve a seamless balance between automated assistance and human expertise. Unlike previous approaches, our method clearly defines the roles and responsibilities of physicians and engineers, aligning with the principle of considering uncertainty about user goals (\cite{10.1145/302979.303030}, Principle 2). Instead of requiring physicians to annotate all images before training, the model first learns independently, and physicians can then use a user-friendly interface in \textit{MEDebiaser} to focus specifically on annotating biased images, thereby reducing the workload associated with full data annotation, and minimizing the costs associated with incorrect predictions (\cite{10.1145/302979.303030}, Principle 8), which replaces the inefficient back-and-forth with a shared interface that enables direct physician annotations and real-time model updates. This approach also empowers physicians to move beyond a passive feedback role; they can actively use familiar annotation methods to correct the model's understanding of biased images, directly incorporating their clinical expertise. \revise{Simultaneously, engineers are repositioned from constant, manual tuning into a supervisory role. While physicians interact with a simplified interface focused on accuracy and visual feedback, engineers can utilize a dedicated metrics dashboard for comprehensive evaluation and validation of the physician-driven updates.} The system integrates their specialized knowledge only when relevant, enabling efficient agent-user collaboration (\cite{10.1145/302979.303030}, Principle 10), thus avoiding unnecessary involvement in the complex modeling process. \revise{This clear division of roles ensures that clinical expertise is effectively integrated while model reliability is maintained, significantly enhancing cross-disciplinary collaboration without overburdening either party.} This targeted involvement not only effectively mitigates bias in MLMIC but also reduces professional barriers and minimizes the need for frequent communication, streamlining physician-engineer collaboration. As a result, both the physical and cognitive workload for physicians and engineers are significantly reduced. In summary, the key contributions of this study are as follows:

\begin{itemize}
\item Through a \textit{formative study}, we analyzed and summarized the collaborative modeling patterns and challenges in MLMIC from the perspectives of both physicians and engineers.
\item \revise{We introduced \textit{MEDebiaser}, an iterative system that redefines physician-engineer collaboration for bias mitigation in MLMIC by providing physicians with an interactive interface to directly address complex technical biases, thereby moving beyond inefficient revision cycles and better integrating expert medical knowledge.}
\item A \textit{mechanism study} and a \textit{user study} demonstrated that \textit{MEDebiaser} significantly reduces bias and optimizes physician-engineer collaboration, confirming its effectiveness and applicability in the medical field.
\end{itemize}

\section{Related Work}
\subsection{Multi-label Classification and Biases}
\label{sec:2.1}
\par Multi-label classification deals with instances that possess multiple labels simultaneously~\cite{Han2022ASO,9568738}. In real-world scenarios, natural image data often exhibits a form of bias~\cite{10.1109/CVPR.2011.5995347,2015arXiv150501257T}. This bias typically stems from low-quality training data, especially noisy labels~\cite{Yuan2021}. Therefore, many approaches focus on addressing this issue at the dataset level~\cite{10368332,10148380}. For instance, Reweighter~\cite{10368332} is a visual analytics tool that mitigates label quality issues through sample reweighting. During model training, bias primarily arises from an \textit{Imbalanced Distribution}~\cite{10.1145/3432931, chouldechova2018frontiersfairnessmachinelearning}, where the training set is highly skewed towards particular labels. Traditional resampling methods are mostly based on a single-label setup~\cite{6871400,10.1016/j.neucom.2015.04.120}.
Additionally, Ridnik et al.~\cite{9710171} adjusted the focal loss functions to create a dynamically balanced training process. Despite these techniques mitigating the problem of imbalanced distribution, AI models still face the issue of \textit{Label Co-occurrence}, where images in the training set frequently feature contextual objects that co-occur with particular labels~\cite{10.1007/978-3-030-01219-9_47}. For example, images containing balls might be incorrectly labeled as dogs because balls and dogs often appear together. This issue is defined as contextual bias in some natural image datasets~\cite{10.1145/3555590}. Several studies employ mathematical or algorithmic approaches to mitigate contextual bias.
With the development of ML, Graph convolutional networks have been widely applied to leverage graph structure information, learning label features within the graph to enhance feature representation, thereby improving classification accuracy and effectiveness~\cite{9010734,10.1007/978-3-030-58589-1_39,8953276,9369105}.

\par In medical image classification, the multi-label bias problem has also garnered considerable attention~\cite{ZHANG2023102772,10.1145/3561653,8961143,10170751}. Unlike natural image datasets, the bias in medical images usually arises less from noisy labels, as they are mostly derived from diagnoses and reports by physicians~\cite{pavlopoulos-etal-2019-survey}. Moreover, the scarcity of medical data makes addressing issues from a dataset perspective more challenging. During training, medical images exhibit similar phenomena to natural image datasets. Due to varying incidence rates of diseases, images of some rare symptoms are difficult to obtain~\cite{ZHANG2023102772}. Consequently, the \textit{Imbalanced Distribution} in medical image datasets is more pronounced and severe, leading to poorer learning outcomes and higher error rates in identifying rare symptoms. In medical imaging, \textit{Label Co-occurrence} becomes more complex than in natural image datasets. The co-occurrence of labels in medical datasets may arise from symptoms that often appear together, such as certain complications~\cite{Wang2024MultilabelLA}. These symptoms may have very similar features in the images and be spatially close, making it difficult for the model to distinguish between different symptoms.
The CXR-LT competition of the ICCV CVAMD 2023 Shared Task~\cite{HOLSTE2024103224} aims to address the issue of long-tail distribution in multi-label thoracic disease classification in chest X-rays, with solutions delivering competitive results that somewhat alleviate the bias in medical multi-label images. In this challenge, Kim et al.~\cite{10350964} proposed a solution called \textit{CheXFusion}, a transformer-based fusion model, which improves classification accuracy by effectively integrating features extracted from multi-view medical images using self-attention and cross-attention mechanisms.

\par Our work leverages pre-trained models for multi-label classification, specifically addressing challenges like \textit{Imbalanced Distribution} and \textit{Label Co-occurrence} in medical image datasets through a human-AI interactive feedback mechanism. Our method can be applied to various attention-based algorithms, enhancing their performance by improving the model's ability to detect complex patterns, increasing interpretability, boosting adaptability, and clinical applicability.

\subsection{Human-AI Collaboration in Medicine}
\par Human-AI collaboration (HAI) refers to the cooperative effort between humans and AI systems to solve specific problems, leveraging each party's strengths with clearly defined roles~\cite{10.1145/3359297}. As HCI continues to evolve, there are currently three main modes of HAI~\cite{10.1145/3544548.3581469}. The first mode is \textit{AI-assisted Decision-making}, where AI acts as an assistant providing recommendations to users, who then combine these suggestions with their prior knowledge to make the final decision. This mode is widely used in decision support systems across various fields, including finance~\cite{10.1016/j.dss.2012.05.048}, operations research~\cite{Gupta2021ArtificialIF}, and healthcare~\cite{Miotto2018,10.1145/3613904.3642883}.
The second mode is \textit{Human-in-the-loop}, where users intervene in the model training process to improve its performance. Interactive machine learning~\cite{10.1145/3185517} exemplifies this approach. Yimam et al.~\cite{10.1007/978-3-319-23344-4_34} used interactive learning to automatically annotate medical documents. \revise{Guo et al.~\cite{10.1007/978-3-319-31753-3_38} proposed an interactive machine-learning approach for automatically grouping medical images, which was notably enhanced by incorporating expert-defined constraints. Moreover, Calisto et al.~\cite{CALISTO2025103444} found that tailoring an AI's communication style to the clinician's experience level in breast cancer diagnosis significantly reduced diagnostic time and errors.}
In \textit{Joint Action}, users and AI work together as a team to achieve a shared goal, focusing on task allocation between the users and AI~\cite{10.1145/3359297,10.1145/3491102.3517449,10.1145/3613904.3642812}. For example,
\textit{HADT}~\cite{doi:10.1137/1.9781611978032.98} uses a reinforcement learning-based strategy to decide human-machine task allocation in dialogue-based disease screening.

\par In the medical field, there is a growing emphasis on leveraging AI to enhance collaboration between physicians and technology across various tasks~\cite{sun2023ai}. However, when using these tools, physicians frequently rely on engineers for troubleshooting and adjustments, as the modeling process can become time-consuming and challenging~\cite{NASARIAN2024102412}. As a result, engineers play a crucial role in this human-AI collaboration, as their mathematical and algorithmic expertise is essential for refining AI models. Recently, in the fields of HCI and VIS, some work has allowed physicians to be directly involved in the modeling process. For example, Wang et al.\cite{10540286} proposed \textit{KMTLabeler}, a tool designed to involve physicians in labeling medical text. Similarly, Ouyang et al.\cite{ouyang2024twophasevisualizationcontinuoushumanai} introduced a two-phase visualization system to facilitate interactive data analysis between physicians and AI. 
While these co-design approaches~\cite{doi:10.1177/20539517211065248} effectively integrate medical expertise with AI, they often involve complex visualization and interactive interfaces and still require physicians to have a certain foundation of ML knowledge, which can lead to steep learning curves for physicians.

\par To address this gap, our system maintains the \textit{Human-in-the-loop} mode, but it aims to redefine the role of AI from a mere tool to a bridge that enhances collaboration between physicians and engineers. We focus on designing straightforward, intuitive, and user-friendly interactions that, through interactive machine learning, enable physicians to engage more directly with the model. This approach not only reduces the workload of human parties but also optimizes the physician-engineer collaboration process.

\subsection{Feedback in Interactive Machine Learning}
\par Interactive machine learning is a paradigm where one or more users iteratively build and refine mathematical models through a cyclical process of input, review, and modification~\cite{10.1007/s12650-018-0531-1}.

\par In interactive machine learning, the AI model processes inputs and generates appropriate outputs based on its understanding of the process that humans aim to represent. The feedback provided by AI to humans is facilitated through explainable AI (XAI). XAI has been applied across various fields to elucidate the inner workings of models, thereby increasing human trust in AI~\cite{A2023100230,CALISTO2022102922,NASARIAN2024102412}. In image recognition tasks, local explanation is widely used in XAI due to its visual intuitiveness~\cite{10295394,10.1117/1.JBO.27.1.015001,9410346}.
In medical imaging, Grad-CAM~\cite{8237336} is employed to visualize the productivity of AI models. Ouyang et al.~\cite{10295394} utilized Grad-CAM to generate saliency maps on MRI images, highlighting important areas due to its effectiveness in dealing with the distinct characteristics of medical images.
In addition to XAI, visual analytics systems are frequently used to deliver AI-analyzed information to humans, exemplified by tools like \textit{OoDAnalyzer}~\cite{10.1109/TVCG.2020.2973258} and \textit{VATLD}~\cite{9233993}. These systems leverage sophisticated interactive visualizations and advanced computational techniques to provide actionable insights, but are primarily designed for users with a background in machine learning or data science, which limits their accessibility to physicians without technical expertise.

\par Humans play a crucial role by contributing their domain expertise related to the data or models in interactive machine learning, thereby guiding model training. The feedback provided by humans, known as human input, is becoming increasingly integrated into AI research as part of human-AI collaboration. This integration primarily occurs through two modes: \textit{Rule-based} and \textit{Attention-guided}~\cite{10.1145/3555590}. The \textit{Rule-based} approach involves embedding predefined expert rules into the model before training to guide its learning.
In medical image analysis, experts annotate prior knowledge beforehand. However, these expert rules often lack generality, and purely \textit{Rule-based} methods may be insufficient for handling complex and dynamic scenarios~\cite{10.1145/3555590}. Additionally, creating these rules requires significant human effort, such as annotating medical datasets, making this approach less practical and user-friendly for ML-naive users~\cite{10.1145/3185517}. \textit{Attention-guided} approaches, on the other hand, offer an effective strategy when embedding principles or rules is not feasible. The attention branch network~\cite{8953929} allows users to directly modify the model's attention on images, facilitating an interactive machine-learning process. \revise{This principle of guiding a model is foundational to the fine-tuning of modern architectures~\cite{ALBAHRI2023156}. \textit{GRADIA}~\cite{10.1145/3555590} utilizes the interactive attention alignment framework to balance prediction accuracy and attention accuracy through human adjustments to model attention.}

\par Current work on MLMIC has not fully leveraged the feedback between humans and AI to mitigate biases arising during model training. Our approach addresses this gap through an interface \textit{MEDebiaser}. \revise{Existing attention-guided systems often require physicians to understand and manipulate an abstract representation of the model's focus, which can be unintuitive~\cite{10.1145/3290605.3300809,Chen2022}.} In contrast, by using local explanations to demystify the model training process and present to physicians, \textit{MEDebiaser} enables them to identify and correct biases directly before resuming training. Unlike traditional systems that often rely on complex and visually striking interactive designs, \textit{MEDebiaser} is built on the principle of using methods that are familiar and intuitive to physicians. This approach prioritizes accessibility and usability, ensuring that the system is both effective and user-friendly for physicians.

\section{Formative Study}
\par The objective of our formative study is to comprehensively explore the perspectives of both physicians and engineers on \textit{Multi-label Medical Image Classification} and \textit{Human-AI Collaboration} tasks, specifically addressing the following research questions: \revise{\textbf{RQ1:} \textit{What are the current collaboration practices between physicians and engineers in building medical models, and what are the opportunities and challenges?} \textbf{RQ2:} \textit{How do physicians and engineers perceive biases in multi-label medical images, and how have such biases been addressed previously together?}}
\par To achieve this, we conducted \textit{Semi-structured Interview} to delve into their challenges and expectations with institutional IRB approval. The insights gained from the interview enabled us to identify six key design challenges (\textbf{C1}-\textbf{C6}) and establish seven design goals (\textbf{DG1}-\textbf{DG7}).

\subsection{Semi-structured Interview}
\subsubsection{Participant} 
\par We recruited five physicians (\textbf{D1}-\textbf{D5}) and four engineers (\textbf{P1}-\textbf{P4}) from local universities and hospitals (mean age = 34.11, SD = 5.23; 5 males and 4 females). Detailed participant information is provided in~\autoref{tab:participant}. All participants possess substantial experience in their respective fields and have collaborated on medical-related machine learning tasks with either physicians or engineers.

\begin{table*}
  \caption{The details of \textbf{Semi-structured Interview} participants.}
  \label{tab:participant}
  \begin{tabular}{ccccccc}
    \toprule
    \textbf{ID} & \textbf{Gender/Age} & \textbf{Research Area}                & \textbf{Experience} & \textbf{Title}               & \textbf{Familiarity with AI} \\
    \midrule
    \textbf{D1} & Male/42  & Cardiothoracic Surgery       & 17 Years                  & Clinical Professor  & Aware               \\
    \textbf{D2} & Female/36  & Cardiothoracic Surgery       & 12 Years                  & M.D.      & Neutral             \\
    \textbf{D3} & Male/33  & Ear, Nose, and Throat        & 7 Years                   & M.D.      & Aware               \\
    \textbf{D4} & Female/31  & Traditional Chinese Medicine & 6 Years                   & Postdoc             & Unfamiliar          \\
    \textbf{D5} & Male/29  & Orthopedic Surgery           & 5 Years                   & Postdoc             & Neutral             \\
    \midrule
    \textbf{P1} & Female/41  & Machine Learning             & 15 Years                  & Associate Professor & Expert              \\
    \textbf{P2} & Male/37  & Machine Learning             & 12 Years                  & Assistant Professor & Expert              \\
    \textbf{P3} & Male/31  & Human-Computer Interaction                          & 7 Years                   & Postdoc             & Familiar            \\
    \textbf{P4} & Female/27  & Data Scientist               & 5 Years                   & Ph.D.              & Expert            \\
    \bottomrule
  \end{tabular}
\end{table*}

\subsubsection{Method} 
\par Before the interview began, each participant signed an informed consent form that addressed privacy, ethics, and data collection for academic purposes. The semi-structured interview was conducted in a focus group format, with 9 participants engaging in the discussion for about 60 minutes. The sessions were moderated by an author, ensuring that each participant had an opportunity to share their perspectives.

\par The focus group discussion centered on the diverse viewpoints of physicians and engineers regarding MLMIC, including the unique characteristics and technical challenges of medical images. Additionally, the interview explored the collaboration modes between physicians and engineers, identifying potential challenges in their interaction. The discussion was audio-recorded with the participant's consent to facilitate accurate analysis later.

\subsubsection{Analysis} 
\par We transcribed the audio recordings into text scripts and corrected transcription errors. Using thematic analysis~\cite{10.1093/acprof:oso/9780199753697.001.0001}, we conducted a comprehensive examination of the interview data. Initially, all authors read through the scripts to achieve a shared understanding, then proceeded with the coding process. During the initial coding phase, two authors segmented the text data into meaningful units and assigned labels to each segment. Subsequently, two other authors refined and adjusted these labels, uncovering additional patterns and themes. Finally, all authors shared and discussed the coding results, reaching a consensus to ensure consistency and reliability. Based on the coding results, we summarized the codebook (\autoref{tab:codebook}), which includes $4$ themes and $13$ labels, with definitions clarified based on the interviewees' perspectives. Through this rigorous process, we addressed \textbf{RQ1} and \textbf{RQ2}, summarizing the challenges and needs faced by physicians and engineers in MLMIC during their collaboration with AI.

\begin{table*}
  \caption{Codebook for \textbf{Semi-structured Interview}, targeting at \textbf{RQ1} and \textbf{RQ2}.}
  \label{tab:codebook}
  \begin{tabular}{llp{9cm}}
    \toprule
    \textbf{Theme} & \textbf{Label} & \textbf{Definition} \\
    \midrule
    Collaboration & Physician's role & Roles and responsibilities of physicians \\
    & Engineer's role & Roles and responsibilities of engineers \\
    & AI Model's role & Roles and responsibilities of AI models \\
    & Collaboration Workflow & Mode, steps, timing, and other details of collaboration \\
    & Collaboration Quality & Challenges, outcomes, feelings, and evaluations of collaboration \\
    \midrule
    Multi-label & Data Quality & Accuracy \& completeness of datasets and distribution \& co-occurrence of labels \\
    & Medical Features & Special characteristics of medical multi-label images \\
    & Potential Biases & Potential biases that may arise from using medical multi-label images \\
    \midrule
    AI Model & Model Performance & Accuracy, validity, and stability of model \\
    & Model Explainability & Clear presentation of model results and decision processes \\
    & Knowledge Integration & Opinions and concerns on integrating human knowledge into model training \\
    \midrule
    Expectation & Interactivity & Interface's ability to facilitate Human-AI collaboration \\
    & Feedback & Clear presentation of model metrics and improvements \\
    \bottomrule
  \end{tabular}
\end{table*}

\subsection{Findings}
\subsubsection{Challenges in Multi-label Medical Image Classification}
\par \textit{Physician's Perspective.} All physicians underscored the unique complexities of multi-label medical images (N = $5$). As \textbf{D2} pointed out, ``\textit{Medical images often have a lot of features of different lesions. Even for us, it takes a high level of expertise and careful attention, so I guess it's a real challenge for AI.}'' Additionally, many physicians highlighted the issue of imbalanced label distribution in medical images (N = $3$). According to \textbf{D1}, ``\textit{...a lot of symptoms are rare, and we might only see a few cases over several months, making it hard to capture them during [data] collection.}'' Furthermore, several physicians discussed the phenomenon of \textit{Label Co-occurrence} (N = $3$). \textbf{D4} commented, ``\textit{Different lesions can be connected or show up together, like certain complications, especially in tongue images, which we might miss if we don't examine [them] carefully.}'' These insights underscore the significance of \textit{Imbalanced Distribution} and \textit{Label Co-occurrence} as prevalent and critical issues in medical imaging. This leads to the first challenge: \textbf{C1. The intricate and multifaceted nature of multi-label medical images.}

\par \textit{Engineer's Perspective.} Engineers believe that multi-label medical images face more severe challenges related to \textit{Imbalanced Distribution} and \textit{Label Co-occurrence} compared to multi-label natural images. They note that existing models and methods designed for multi-label classification may not perform well on medical datasets. \textbf{P4} observed, ``\textit{The data from physicians usually has a really obvious long-tail [distribution]. Plus, a lot of the features in the images are not only very similar but also close together, and they might even overlap.}'' \textbf{P1} added, ``\textit{For these medical images, we usually hope that physicians can annotate them, but they're often not willing to spend a lot of time and effort on it.}'' \textbf{P4} further commented, ``\textit{...if we could analyze the images at the pixel level, the model might pick up on the features better.}'' This leads to the second challenge: \textbf{C2. Insufficient capability for fine-grained learning of the images.}

\subsubsection{Challenges in Human-AI Collaboration}
\par Both physicians and engineers acknowledge the benefits of incorporating human knowledge into the model training process, yet they recognize significant challenges (N = $8$). Physicians highlighted that differences in terminology and understanding create communication barriers with engineers, a concern frequently supported by previous studies~\cite{MARTINNOGUEROL2021317}, which ultimately impacts the model's performance. \textbf{D1} explained, ``\textit{When I talk with collaborators, I try to simplify the medical terms, but they still often don't understand or misinterpret what I'm saying, which means we don't always get the AI results we're aiming for.}'' Engineers echoed this sentiment, with \textbf{P2} stating, ``\textit{...these terms are new to us, so it's harder to grasp. We often struggle to fully understand what the physicians mean, which ends up affecting the functionality or outcomes they're looking for.}'' \textbf{P3} further noted, ``\textit{Sometimes, a lot of the medical concepts or rules that physicians suggest are tough to translate into mathematical models during the process.}'' Consequently, the medical knowledge or rules that physicians aim to integrate into the model are not always effectively communicated or implemented by engineers. This highlights the third challenge: \textbf{C3. The ineffectiveness and difficulty of integrating human knowledge into model training.}

\par Several physicians have reported a decrease in their trust in AI during collaboration (N = $3$). This issue largely stems from the lack of model interpretability, a concern also highlighted in previous research~\cite{A2023100230}. As \textbf{D3} remarked, ``\textit{I don't get how the model comes to these conclusions, and sometimes I can't be sure its decisions match our clinical judgment.}'' Additionally, mistrust can arise from physicians not being involved in the model training and decision-making processes. \textbf{D2} pointed out, ``\textit{Usually, I just get the results from the model, but if I spot [any] errors, I have no way to fix them. This really hurts my trust in the system.}'' Consequently, the collaboration suffers from inadequate feedback between the parties, which significantly erodes the physicians' trust. The fourth challenge is: \textbf{C4. Lack of mutual feedback in the collaboration between physicians and AI models.}

\par When discussing the collaboration between physicians and engineers, both parties raised several specific and nuanced challenges. One major concern was the difference in their working styles (N = $6$). \textbf{D4} emphasized, ``\textit{Our work is usually driven by clinical needs and urgency, while they're more focused on project deadlines and data availability. This difference often causes mismatched timelines and inefficiency.}'' \textbf{P1} added, ``\textit{Physicians' time is incredibly valuable, and they want to see results quickly, but developing and training our models takes time and constant adjustments.}'' In addition, \textbf{D5} mentioned, ``\textit{Previous collaborators gave us some interfaces to help with data analysis or using AI, but I found them hard to use, and some features were tough to understand even after training.}'' The complexity of existing workflows often demands significant time and effort from physicians to learn and engage in the modeling process, underscoring the fifth challenge: \textbf{C5. Inconsistency and high learning costs in the collaboration between physicians and engineers.}

\par Furthermore, engineers pointed out that physicians might experience issues related to the ``overuse'' of AI models (N = $2$). \textbf{P1} observed, ``\textit{In a previous project, there was a case where the model picked up incorrect information because physicians used it improperly, so it's necessary for us to re-evaluate the model's usability.}'' Similarly, \textbf{D5} mentioned, ``\textit{I'm concerned I might make mistakes without even realizing it, so I prefer to double-check the performance with engineers.}'' This highlights the sixth challenge: \textbf{C6. Lack of joint review of the model's feedback by physicians and engineers.}

\subsection{Design Goals}
\par After interviewing both physicians and engineers, we have summarized a series of design goals aimed at effectively addressing the aforementioned challenges. Our approach focuses on improving the traditional collaboration mode between these two parties (\autoref{fig:traditional}), enhancing conventional classification methods to foster more direct feedback and interaction between physicians and AI models. We have organized the seven design goals into two key areas: \raisebox{-0.13cm}{\includegraphics[height=0.45cm]{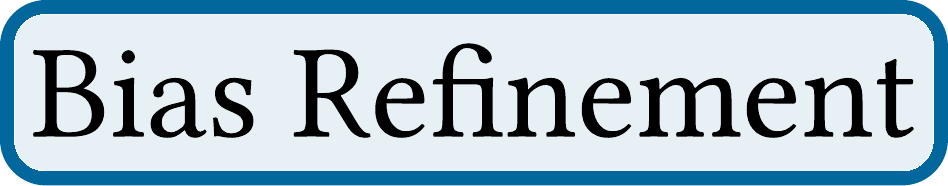}}, which involves \textit{Pixel-level Bias Refinement Integrated with Physicians' Knowledge}, and \raisebox{-0.13cm}{\includegraphics[height=0.45cm]{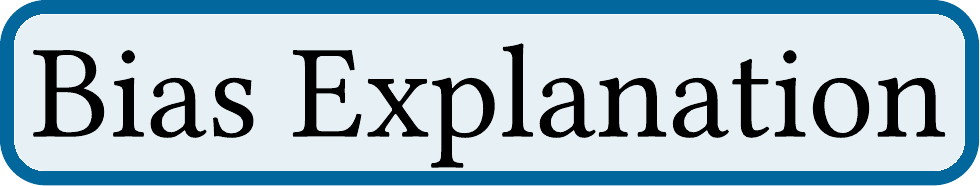}}, which focuses on the \textit{Display and Analysis of Biases During AI Training and Fine-tuning Results}. These components function within an \raisebox{-0.13cm}{\includegraphics[height=0.45cm]{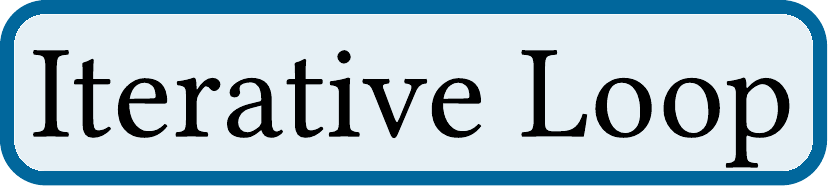}}, continuously reducing bias, refining the model's predictive performance, increasing physicians' engagement and trust, and simultaneously lightening the workload of engineers.

\par\raisebox{-0.13cm}{\includegraphics[height=0.45cm]{Figure/biasrefinement.pdf}} Given the differences in professional expertise, work modes, and technical barriers between physicians and engineers, efforts to incorporate medical knowledge into AI training may not always produce optimal results, leading to inefficiencies in collaboration. However, this integration is essential for effective medical image training. To overcome these challenges (\textbf{C3}, \textbf{C5}), we aim to enhance more direct cooperation between physicians and AI models, while also improving physician-engineer collaboration. This brings us to our first design goal: \textbf{DG1. Facilitate more direct interaction and knowledge integration between physicians and AI models.} This objective was emphasized repeatedly during our semi-structured interview.

\par \raisebox{-0.13cm}{\includegraphics[height=0.45cm]{Figure/biasexplanation.pdf}} To address the fourth challenge (\textbf{C4}), we should not only facilitate the integration of physicians' knowledge into AI but also provide meaningful feedback from the AI back to the physicians. This requires clear displays of the model's training and decision-making processes, enabling physicians to identify biases or inaccuracies generated by the model. For multi-label medical images, it is essential to present the different symptoms within a single image. As \textbf{D3} emphasized during the semi-structured interview, ``\textit{I want the model to clearly mark or highlight the locations of each symptom it identifies. I will only trust it when its judgments align with mine.}'' This underscores the importance of our second design goal: \textbf{DG2. Provide detailed displays of the model's decision-making process during training.}

\par \raisebox{-0.13cm}{\includegraphics[height=0.45cm]{Figure/biasexplanation.pdf}} Beyond providing feedback during AI training, physicians also emphasized the importance of displaying the outcomes of the training process. As \textbf{D2} mentioned, ``\textit{I want to see what changes occur after I correct errors and whether these changes improve the model's performance.}'' Illustrating how physicians' interventions affect model accuracy and reduce bias allows them to better understand the impact of their contributions, which in turn helps validate the model's usability. This brings us to our third design goal: \textbf{DG3. Display the impact of physicians' interventions on model outcomes.} This goal further supports the feedback mechanisms outlined in \textbf{C4} and \textbf{C6}.

\par \raisebox{-0.13cm}{\includegraphics[height=0.45cm]{Figure/biasexplanation.pdf}} While physicians play a crucial role in verifying the model, the expertise of engineers in evaluating and controlling the model is equally essential. Even when physicians find the model's interpretability satisfactory, risks such as overfitting\footnote{Overfitting happens when training data details are learned too specifically, so performance drops on new data.} or gradient vanishing\footnote{Gradient vanishing occurs when gradients become very small, slowing or stopping learning.} could still arise due to excessive bias refinement or inherent model issues. Identifying and addressing these potential problems to keep the model usable requires engineers' specialized knowledge. Therefore, it is vital for both physicians and engineers to collaboratively validate the model. To address \textbf{C6} and supplement \textbf{DG3}, our fourth design goal is: \textbf{DG4. Provide feedback on model performance to engineers.}

\par \raisebox{-0.13cm}{\includegraphics[height=0.45cm]{Figure/biasrefinement.pdf}} To mitigate issues of \textit{Imbalanced Distribution} and \textit{Label Co-occurrence} in MLMIC (\textbf{C1}), engineers typically rely on mathematical and algorithmic techniques. As \textbf{P2} noted, ``\textit{The model's inaccuracies often stem from focusing on incorrect areas.}'' By enabling physicians to correct biases that arise during model training before optimization is passed on to engineers, the model's performance and effectiveness can be significantly improved. However, for ML-naive users, steering AI models can be challenging, as highlighted in \textbf{C5}. Therefore, it's essential to first provide physicians with an understanding of the data and then offer intuitive, user-friendly tools that enable them to easily identify and correct biases. Our fifth design goal is: \textbf{DG5. Provide physicians with data cues and accessible tools to adjust and correct biases during model training.}

\par \raisebox{-0.13cm}{\includegraphics[height=0.45cm]{Figure/biasrefinement.pdf}} The similarity of features and the spatial proximity or overlap in medical images pose significant challenges for effective training. Engineers have observed that for a model to accurately learn the characteristics of various symptoms, it requires more detailed image analysis (\textbf{C2}). However, the annotations needed for this level of detail depend heavily on the expertise of physicians, and such annotated data is often lacking. To address this, we can involve physicians in the annotation process during training, allowing them to highlight key features in the images where detailed learning is required. Thus, our sixth design goal is: \textbf{DG6. Incorporate techniques that enable the model to discern and learn subtle features in medical images.}

\par \raisebox{-0.13cm}{\includegraphics[height=0.45cm]{Figure/iterativeloop.pdf}} Our final design goal is: \textbf{DG7. Develop an interactive interface that facilitates iterative feedback between physicians and AI models.} This goal is crafted to support feedback exchange (\textbf{C4}) and foster more direct and closer interactions (\textbf{C3}) between physicians and AI models. Within this iterative loop, the process of physicians providing feedback to the AI and receiving feedback from the AI is seamlessly integrated. Through continuous interaction, physicians can incrementally reduce model biases and enhance their performance. The importance of this goal lies in its potential to gradually increase physicians' familiarity with human-AI collaboration, allowing them to progressively refine biases and improve the model's performance.

\subsection{Content Analysis}
\par To develop a user-friendly and accessible annotation method (\textbf{DG5}), we first conducted a comprehensive review of the annotation tools commonly employed by our target users. This review included mainstream annotation tools identified through an extensive search across online platforms such as YouTube and GitHub, recent peer-reviewed literature, and interviews with medical imaging experts. As a result, we identified $5$ annotation methods frequently employed in medical AI annotation workflows. We then systematically analyzed their strengths and limitations (\autoref{tab:annotation_methods}, Appendix~\ref{sec:content}), evaluating them based on usability, annotation granularity, and their suitability for medical applications.

\par To empirically evaluate these methods, we designed and conducted a usability study in which participants annotated a sample dataset using each of the identified techniques. The results, summarized in \autoref{tab:usability_results},~\autoref{sec:content}, include both quantitative performance metrics and qualitative feedback. Based on the study's findings and expert consultations, we identified polygon annotation as the optimal method, striking a balance between annotation accuracy and usability for physicians.

\section{MEDebiaser}

\begin{figure*}[h]
\centering
\includegraphics[width=1\linewidth]{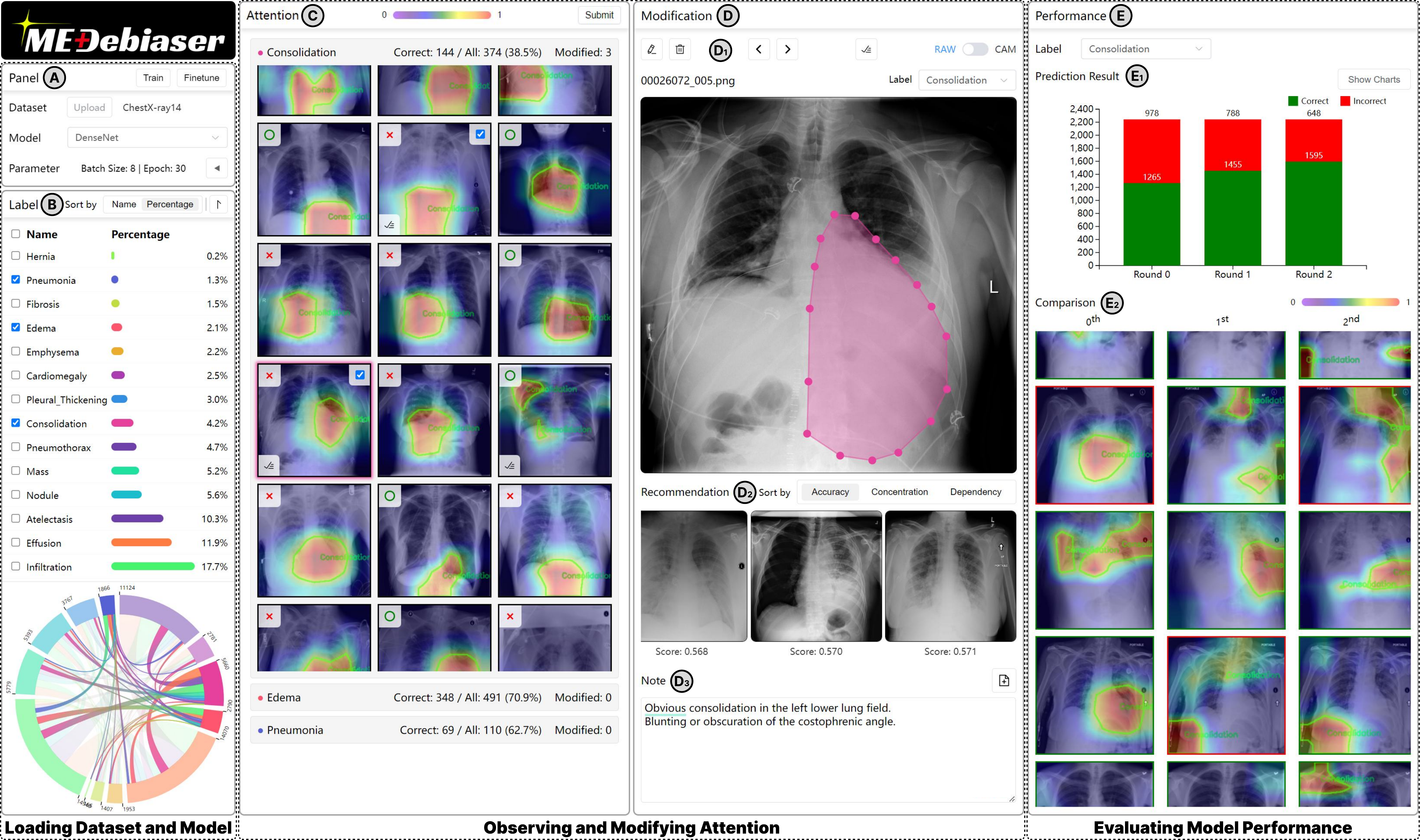}
\caption{\raisebox{-0.13cm}{\includegraphics[height=0.45cm]{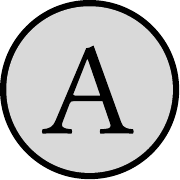}} The \textit{Panel View} provides components for uploading datasets, selecting models, and setting training parameters. \raisebox{-0.13cm}{\includegraphics[height=0.45cm]{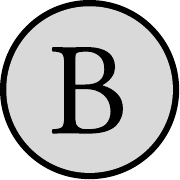}} The \textit{Label View} includes a table displaying label distribution and a chord diagram showing co-occurrence. \raisebox{-0.13cm}{\includegraphics[height=0.45cm]{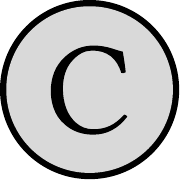}} The \textit{Attention View} displays local explanations for the selected labels. \raisebox{-0.13cm}{\includegraphics[height=0.45cm]{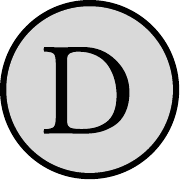}} The \textit{Modification View} includes \raisebox{-0.13cm}{\includegraphics[height=0.45cm]{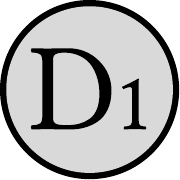}} an \textit{Editing Area} for fine-tuning attention, \raisebox{-0.13cm}{\includegraphics[height=0.45cm]{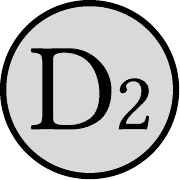}} a \textit{Recommendation Area} for sorting based on metrics, and \raisebox{-0.13cm}{\includegraphics[height=0.45cm]{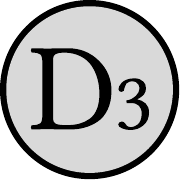}} a \textit{Note Area} for record-keeping. \raisebox{-0.13cm}{\includegraphics[height=0.45cm]{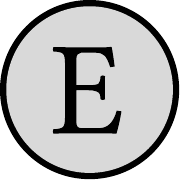}} The \textit{Performance View} contains \raisebox{-0.13cm}{\includegraphics[height=0.45cm]{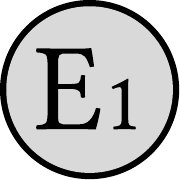}} a \textit{Metrics Area} to show the model's metrics for each round of fine-tuning on specific labels, as well as \raisebox{-0.13cm}{\includegraphics[height=0.45cm]{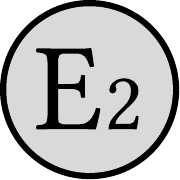}} a \textit{Comparison Area} to display comparisons of local explanations after each round of fine-tuning.}
\label{fig:teaser}
\end{figure*}

\par Based on the identified user needs, we designed and implemented \textit{MEDebiaser} (\autoref{fig:teaser}), an interactive system that enhances collaboration between physicians and engineers in MLMIC, addressing the following research question: \textbf{RQ3:} \textit{What practices do physicians and engineers hope to adopt to mitigate multi-label issues and enhance their collaboration?}

\par
\begin{figure*}[h]
\centering
\includegraphics[width=1\linewidth]{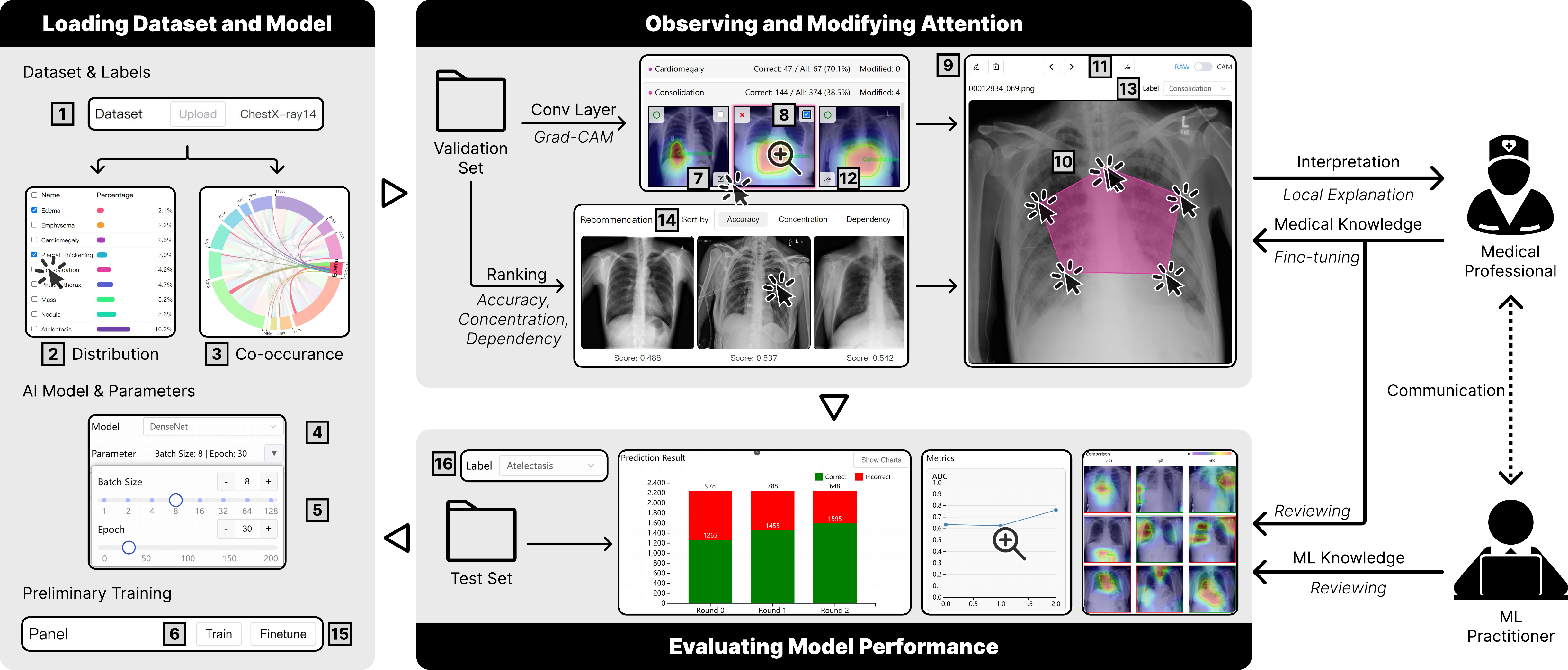}
\caption{The \textit{MEDebiaser} workflow includes three main stages: \textit{Loading Dataset and Model}, \textit{Observing and Modifying Attention}, and \textit{Evaluating Model Performance}.}
\label{fig:pipeline2}
\end{figure*}

 \textit{MEDebiaser} features five main components: the \textit{Panel View} (\autoref{fig:teaser}-\raisebox{-0.13cm}{\includegraphics[height=0.45cm]{Figure/icona.pdf}}) allows users to upload datasets, select models, and adjust training parameters (\textbf{DG1}); the \textit{Label View} (\autoref{fig:teaser}-\raisebox{-0.13cm}{\includegraphics[height=0.45cm]{Figure/iconb.pdf}}) displays the distribution of labels in a table and their co-occurrence (\textbf{DG5}); the \textit{Attention View} (\autoref{fig:teaser}-\raisebox{-0.13cm}{\includegraphics[height=0.45cm]{Figure/iconc.pdf}}, \hspace{1pt}\raisebox{-0.13cm}{\includegraphics[height=0.45cm]{Figure/biasexplanation.pdf}}\hspace{1pt}) provides local explanations during the training and fine-tuning process (\textbf{DG2}, \textbf{DG7}); the \textit{Modification View} (\autoref{fig:teaser}-\raisebox{-0.13cm}{\includegraphics[height=0.45cm]{Figure/icond.pdf}}, \hspace{1pt}\raisebox{-0.13cm}{\includegraphics[height=0.45cm]{Figure/biasrefinement.pdf}}\hspace{1pt}) offers intuitive tools for pixel-level fine-tuning (\textbf{DG1}, \textbf{DG5}, \textbf{DG6}, \textbf{DG7}); and the \textit{Performance View} (\autoref{fig:teaser}-\raisebox{-0.13cm}{\includegraphics[height=0.45cm]{Figure/icone.pdf}}, \hspace{1pt}\raisebox{-0.13cm}{\includegraphics[height=0.45cm]{Figure/biasexplanation.pdf}}\hspace{1pt}) displays prediction results and the impacts of user interventions on model outcomes (\textbf{DG3}, \textbf{DG4}). Users can then initiate a new round of fine-tuning based on the retrained model, following an \raisebox{-0.13cm}{\includegraphics[height=0.45cm]{Figure/iterativeloop.pdf}} process. The \textit{MEDebiaser} workflow (\autoref{fig:pipeline2}) is structured into three main stages: \textit{Loading Dataset and Model}, \textit{Observing and Modifying Attention}, and \textit{Evaluating Model Performance}. By iteratively executing these stages, \textit{MEDebiaser} enables physicians to continuously correct biases that arise during model training, seamlessly incorporating their medical expertise into the training process. To be noted, the actual application of the system is not restricted to the specific dataset and model showcased.

\subsection{Loading Dataset and Model}

\par At this stage, the users can upload datasets, review label distribution and co-occurrence within the dataset, select the model to be used, and configure the appropriate training parameters.

\subsubsection{Dataset}

\par To evaluate the system, we conducted tests on several datasets, with \textbf{ChestX-ray14}~\cite{8099852} being one of the most representative examples. This dataset includes $112,120$ frontal view chest X-ray images, each at a resolution of $1024\times1024$, collected from $30,805$ patients between $1992$ and $2015$. These images have been automatically labeled into $14$ different thoracic pathology categories based on radiological diagnostic reports, with each image potentially containing multiple labels. As shown in~\autoref{fig:distribution}, Appendix~\ref{sec:Dataset}, the label distribution is highly imbalanced, with the most frequent label appearing nearly $20,000$ times and the least frequent only $227$ times. Additionally, there is a noticeable co-occurrence among the labels, as shown in~\autoref{tab:Matrix}, Appendix~\ref{sec:Dataset}.

\subsubsection{Model}
\par \revise{To tackle the unique challenges of MLMIC, \textit{MEDebiaser} employs a one-vs-rest classification strategy, treating the problem as N parallel binary classification tasks, where N is the number of possible labels. Architecturally, the system utilizes a powerful convolutional neural network (CNN) backbone---such as DenseNet~\cite{8099726} or ResNet~\cite{7780459}---which outputs an N-dimensional logit vector. Each logit is then passed through a sigmoid activation function, rather than the mutually exclusive softmax, to independently compute each class's probability.} For training, the model minimizes the mean of N binary cross-entropy (BCE) losses against the multi-hot ground truth vector. This entire approach is inherently robust for handling common issues like \textit{Imbalanced Distribution} and \textit{Label Co-occurrence}. To ensure strong initial performance, the selected CNN models are pre-trained on extensive datasets like ImageNet~\cite{5206848} before training commences on the user's dataset.

\subsubsection{Interaction}
\par Initially, users can upload their datasets through the \textit{Panel View} by clicking \raisebox{-0.17cm}{\includegraphics[height=0.5cm]{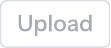}} (\autoref{fig:pipeline2}-\raisebox{-0.07cm}{\includegraphics[height=0.35cm]{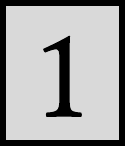}}). Once the dataset is uploaded, users can observe the proportion and distribution of each label in the \textit{Label View} via a table (\autoref{fig:pipeline2}-\raisebox{-0.07cm}{\includegraphics[height=0.35cm]{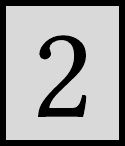}}). The sorting feature allows users to easily identify labels with smaller proportions, which may be more prone to biases and errors during model training and thus require special attention (\textbf{DG5}). Each label is color-coded in the table, and in the accompanying chord diagram (\autoref{fig:pipeline2}-\raisebox{-0.07cm}{\includegraphics[height=0.35cm]{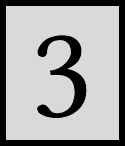}}), which is well-suited for showing data with associative relationships within the dataset~\cite{10.1145/3678698.3678707}. Each arc connects to its co-occurring labels, with the band thickness representing the frequency of co-occurrence. This diagram enables users to clearly observe label co-occurrence and identify features that may need additional focus (\textbf{DG5}). Next, users can select a model from a predefined set using \raisebox{-0.17cm}{\includegraphics[height=0.5cm]{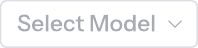}} (\autoref{fig:pipeline2}-\raisebox{-0.07cm}{\includegraphics[height=0.35cm]{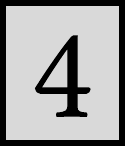}}). To ensure the model can run on various machines, we provide commonly used CNN models for multi-label scenarios with default settings (batch size = $4$, epoch = $30$), \revise{which simplifies the technical setup and allows physicians to proceed with confidence without deep ML knowledge.} These default parameters are highly versatile and compatible with different hardware environments, computational capabilities, and workflows in healthcare facilities. For physicians who need to modify parameters or are familiar with ML, clicking on \raisebox{-0.17cm}{\includegraphics[height=0.5cm]{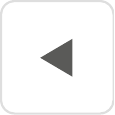}} allows direct adjustment of these parameters and models using \raisebox{-0.17cm}{\includegraphics[height=0.5cm]{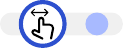}} (\autoref{fig:pipeline2}-\raisebox{-0.07cm}{\includegraphics[height=0.35cm]{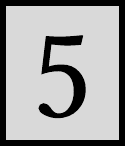}}). This flexibility offers more diverse options to optimize training according to their specific needs, training time, and hardware requirements. The user then clicks \raisebox{-0.17cm}{\includegraphics[height=0.5cm]{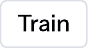}} (\autoref{fig:pipeline2}-\raisebox{-0.07cm}{\includegraphics[height=0.35cm]{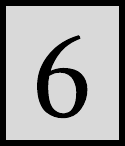}}), and the preliminary training begins. Upon completion, users can review the performance of each label in the \textit{Performance View}. Labels that require further attention can be selected by clicking \raisebox{-0.07cm}{\includegraphics[height=0.3cm]{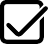}} in the table or by clicking on the arcs in the chord diagram, which will then highlight them in the \textit{Attention View}.

\subsection{Observing and Modifying Attention}
\par This stage is designed to achieve two main objectives: first, to present the model's decision-making process to the physicians (\textbf{DG2}); and second, to enable physicians to refine the model's attention on biased images by incorporating human expertise (\textbf{DG1}). \revise{This is accomplished through an iterative ``observe-annotate-fine-tune'' active learning loop. To address the first objective, the system provides \textit{local explanation} by displaying Grad-CAMs on validation images. To streamline the review process, a \textit{customized ranking strategy} then prioritizes the most problematic cases based on three active learning criteria: prediction accuracy, attention concentration, and co-occurrence dependency. Guided by this \textit{ranking}, physicians can achieve the second objective by annotating correct regions using pixel-level masks. The model is subsequently \textit{fine-tuned} with a dynamically weighted joint prediction. Importantly, Grad-CAM is leveraged here not just for interpretation, but for rapid error localization and correction, improving model performance with minimal expert effort while enhancing collaboration.}

\subsubsection{Local Explanation}
\par During the preliminary training phase, models trained on the \textbf{ChestX-ray14} dataset are used to make predictions on the validation subset. Simultaneously, Grad-CAM is applied to the last convolutional layer of the CNN model, offering an effective XAI technique for interpreting CNNs~\cite{8237336}. Unlike traditional interpretability methods such as SHAP~\cite{10.5555/3295222.3295230} and LIME~\cite{ribeiro-etal-2016-trust}, which can struggle with feature correlations or focus on small regions of the decision space, Grad-CAM leverages the natural structure of CNNs to highlight the most influential image regions~\cite{8237336}. This results in higher efficiency and better integration into rapid-feedback workflows. While Guided Backpropagation~\cite{springenberg2015strivingsimplicityconvolutionalnet}, a commonly used CNN visualization method, emphasizes pixel-level contributions, often amplifying noise in multi-label scenarios, Grad-CAM mitigates this by utilizing higher-level feature maps, offering clearer and more intuitive explanations for physicians. Specifically, Grad-CAM calculates the gradients of a target label with respect to the feature maps of the selected convolutional layer. These gradients are then globally averaged to determine the importance weights for each feature map, highlighting the regions most influential in the model's predictions. The weighted combination of feature maps is visualized as a heatmap, overlaid on the original image to indicate the spatial areas crucial to the model's decision-making process. A comparison of the methods discussed above is provided in~\autoref{fig:visualization}, ~\autoref{sec:attention}.

\subsubsection{Fine-tuning}
\label{sec:4.2.2}
\par \revise{In the modification phase of \textit{MEDebiaser}'s workflow, physicians critically assess the model's heatmaps and provide pixel-level feedback by annotating incorrect attention areas with polygon masks. To incorporate this clinical feedback, we fine-tune the model using a dynamically weighted joint loss function. This function is composed of two distinct components: for the prediction loss, we adopt the function from the Explanation-guided Learning framework~\cite{gao2024going}, an approach whose effectiveness is supported by studies like MAGI~\cite{zhang2023magi}. The attention loss, in contrast, is defined as the mean squared error between the model's Grad-CAM heatmap and the physician-provided mask, which helps the model focus on important features, improving accuracy. Crucially, the balance between these two losses is dynamically adjusted, with the weights for the attention loss being proportionally tailored based on the frequency of each label in the dataset. This combined approach ensures a balanced emphasis during training and effectively steers the model's focus toward the correctly annotated regions to optimize both prediction accuracy and explanation clarity.}

\subsubsection{Ranking}
\par In the proposed iterative refinement process for \textit{MEDebiaser}, active learning~\cite{10.1145/3472291} is leveraged to enhance the efficiency of image annotation. The process begins by ranking images based on their likelihood of errors, with the model prioritizing those it finds most uncertain. This ranking strategy, a core principle of active learning, directs physicians’ attention to the most critical cases---where the model's predictions are least certain or most prone to error. By focusing on these uncertain samples, physicians provide high-value annotations that significantly contribute to model improvement. This approach streamlines the identification and correction of potential inaccuracies early in the refinement cycle, enabling the model to learn from the most informative data and progressively enhance its accuracy with minimal human effort.

\par \textit{Iterative Refinement Cycle}. The workflow follows a cyclical process where physicians start by annotating images predicted to be most susceptible to errors. The model is then fine-tuned with this new data, leading to the generation of updated heatmaps for the remaining images. These images are re-ranked based on the revised error likelihood assessments. This cycle continues, with each iteration progressively enhancing the model's accuracy and interpretability until the physicians determine that the model has reached a satisfactory level of performance.

\par \textit{Customized Ranking Strategy}. To streamline the process, a three-tiered ranking system is implemented, tailored to align with the preferences and expertise of physicians:
\begin{itemize}
\item \textbf{Label Prediction Accuracy:} Experts suggest prioritizing images where the model's confidence significantly deviates from certainty. These images are ranked based on the deviation of their predictive values from the ideal, with values close to 1 indicating high accuracy and confidence, and values close to 0 representing low confidence and a higher likelihood of misclassification.

\item \textbf{Heatmap Concentration:} The system analyzes the model's attention focus by examining the heatmap's $n\times n$ matrix, where $n$ represents the image's pixel dimensions. \revise{To ensure a fair and comparable analysis across different images, each Grad-CAM heatmap is first normalized using global min-max scaling, which rescales all values to a range between 0 and 1. This normalization is critical as it allows for ranking based on relative attention deviation rather than absolute, unscaled gradient magnitudes.} A value approaching 1 indicates a strong, concentrated focus on the most critical region, while a lower value suggests the model's attention is more diffuse.

\item  \textbf{Co-occurrence Matrix Dependency:} Experts believe that labels that frequently co-occur with other labels are more likely to be misclassified. To quantify this, the dependency between labels is evaluated using a co-occurrence matrix $M$, where $M_{ij}$ represents the frequency of co-occurrence between label $i$ and label $j$. \revise{To improve the discriminative power of this metric, we use inverse frequency instead of direct frequency, which reduces the dominance of ubiquitous labels and highlights more meaningful, strongly associated pairs.} The inverse frequency for a target label $c$ with each label $j$ is computed as~\eqref{equ:freq}:

\begin{equation}
    \label{equ:freq}
  \text{inverse frequency}(c, j) = \frac{1}{M_{cj} + 0.01}
\end{equation}

\par The small constant \( 0.01 \) ensures stability by preventing division by zero. The overall dependency score for the target label \( c \) is then obtained by normalizing these inverse frequencies and summing over the relevant labels. \revise{To prevent rarely co-occurring labels from inflating the scores, we include only ``\( \text{positive labels} \)''---those whose co-occurrence with \( c \) exceeds a defined threshold---ensuring that weak or incidental associations have no impact on the final score. The dependency score is calculated as~\eqref{equ:dependency}:}

\begin{equation}
\label{equ:dependency}
  \text{dependency score}_c = \sum_{j \in \text{positive labels}} \frac{\text{inverse frequency}(c, j)}{\sum_{k} \text{inverse frequency}(c, k)}
\end{equation}

\par A dependency score close to $1$ indicates a high likelihood of misclassification due to strong interdependencies, while a score close to $0$ suggests lower interdependency and thus fewer challenges in classification.

\end{itemize}

\subsubsection{Interaction}
\par For the first objective, the \textit{Attention View} provides a clear display of the Grad-CAM visualization of the model's decision-making process on the validation set. Users can zoom in on any image by clicking on it, with icons on the top left indicating the accuracy of the model's label recognition--- \raisebox{-0.06cm}{\includegraphics[height=0.3cm]{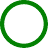}} for correct predictions and \raisebox{-0.07cm}{\includegraphics[height=0.3cm]{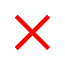}} for incorrect ones. This allows physicians to assess whether the local explanations for the selected labels are reasonable, particularly for symptoms that are less frequent or often co-occur with others (\textbf{DG2}). If an explanation is deemed unreasonable, users can click \raisebox{-0.15cm}{\includegraphics[height=0.5cm]{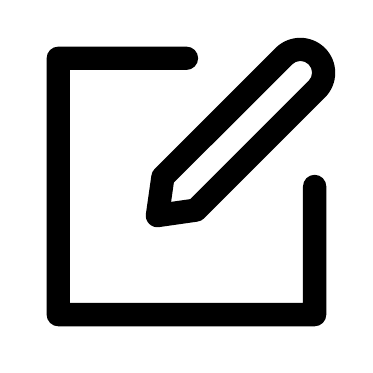}} (\autoref{fig:pipeline2}-\raisebox{-0.07cm}{\includegraphics[height=0.35cm]{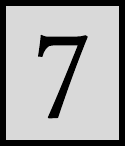}}) in the bottom right, which will bring the corresponding image into the \textit{Modification View}. For those reasonably predicted, users can click \raisebox{-0.07cm}{\includegraphics[height=0.3cm]{Figure/checkbox.pdf}} (\autoref{fig:pipeline2}-\raisebox{-0.07cm}{\includegraphics[height=0.35cm]{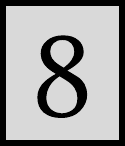}}) located in the top right corner. This action will save the Grad-CAM of the selected image as polygons, which will then be used in the next round of fine-tuning.

\par For the second objective, we employ annotation methods familiar to physicians~\cite{10.5555/3122929}, allowing them to mark areas of interest directly on the image using polygons, \revise{which is an intuitive process that does not require ML knowledge, thereby lowering the technical barrier for them to contribute their domain expertise (\textbf{DG5}).} In the \textit{Modification View}, users can enter annotation mode by clicking \raisebox{-0.17cm}{\includegraphics[height=0.5cm]{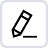}} (\autoref{fig:pipeline2}-\raisebox{-0.07cm}{\includegraphics[height=0.35cm]{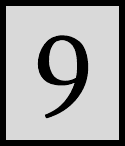}}). They can then draw polygons by clicking on the image and double-clicking to finish the annotation (\autoref{fig:pipeline2}-\raisebox{-0.07cm}{\includegraphics[height=0.35cm]{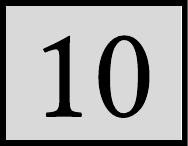}}). Annotating at the pixel level enables the model to focus on more detailed features (\textbf{DG6}). After completing the annotation, users can save their work by clicking \raisebox{-0.17cm}{\includegraphics[height=0.5cm]{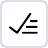}} (\autoref{fig:pipeline2}-\raisebox{-0.07cm}{\includegraphics[height=0.35cm]{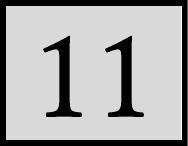}}), storing the polygon for the next round of fine-tuning. Annotated images will display \raisebox{-0.15cm}{\includegraphics[height=0.5cm]{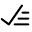}} (\autoref{fig:pipeline2}-\raisebox{-0.07cm}{\includegraphics[height=0.35cm]{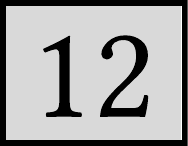}}) in the bottom left corner of the images in \textit{Attention View}. Users can also differentiate between various labels within the same image by selecting from \raisebox{-0.17cm}{\includegraphics[height=0.5cm]{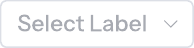}} (\autoref{fig:pipeline2}-\raisebox{-0.07cm}{\includegraphics[height=0.35cm]{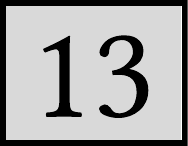}}) and annotating different pixel-level areas accordingly.

\par In the \textit{Recommendation Area} (\autoref{fig:teaser}-\raisebox{-0.13cm}{\includegraphics[height=0.45cm]{Figure/icond2.pdf}}), we prioritize images requiring attention by calculating label prediction accuracy, heatmap concentration, and co-occurrence matrix dependency. Users can choose one of these three methods to sort and focus on the images most in need of modification (\autoref{fig:pipeline2}-\raisebox{-0.07cm}{\includegraphics[height=0.35cm]{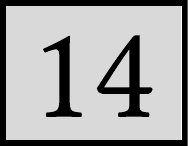}}). Additionally, in the \textit{Note Area} (\autoref{fig:teaser}-\raisebox{-0.13cm}{\includegraphics[height=0.45cm]{Figure/icond3.pdf}}), users can take notes on the image. Compared to the traditional practice of annotating entire datasets, these designs significantly reduce the workload of physicians. 

\par After labeling the biased images and saving the polygon information, users can click \raisebox{-0.17cm}{\includegraphics[height=0.5cm]{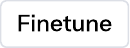}} (\autoref{fig:pipeline2}-\raisebox{-0.07cm}{\includegraphics[height=0.35cm]{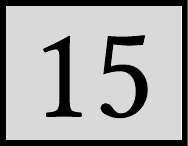}}) to start a new round of fine-tuning (\textbf{DG7}). Once completed, the results will be updated in the \textit{Performance View}.

\subsection{Evaluating Model Performance}
\par In this stage, users evaluate the model's effectiveness by analyzing parameters and Grad-CAM on the test set.

\par In the \textit{Performance View}, users begin by selecting the label they wish to review from \raisebox{-0.17cm}{\includegraphics[height=0.5cm]{Figure/select_label.pdf}} (\autoref{fig:pipeline2}-\raisebox{-0.07cm}{\includegraphics[height=0.35cm]{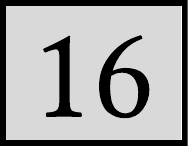}}). The \textit{Prediction Results} (\autoref{fig:teaser}-\raisebox{-0.13cm}{\includegraphics[height=0.45cm]{Figure/icone1.pdf}}) displays, for each round of fine-tuning, the number of correctly and incorrectly predicted labels on the test set. This visualization enables users to track the trend in correct predictions over time, providing insight into model performance progression. By observing the trends in these charts, physicians can assess whether their modifications are steering the model in a positive direction (\textbf{DG3}). Clicking on the \raisebox{-0.17cm}{\includegraphics[height=0.5cm]{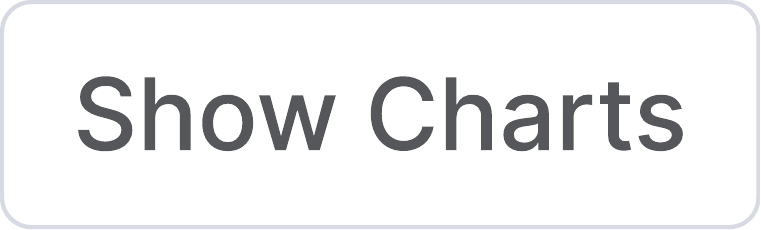}} button reveals the evaluation metrics for each round, including \textit{precision}, \textit{recall} (or \textit{sensitivity}), \textit{F1 score}, and \textit{AUC}. The rationale for selecting these specific metrics—while omitting others—is detailed in \autoref{tab:metrics},~\autoref{sec:Metrics}. To accommodate the varying levels of technical expertise among users, the system adopts a layered presentation strategy. While these parameters are not directly shown in the default view, they remain accessible through interaction for users with machine learning knowledge, such as engineers, to verify that the model remains functional and reliable (\textbf{DG4}). At the same time, physicians are provided with visual trends that reflect performance changes in a more intuitive and accessible manner, aligned with their diagnostic thinking patterns. This design intentionally abstracts away complex terminology, reducing cognitive load for non-expert users, while preserving access to detailed metrics when needed. In doing so, the system strikes a balance between interpretability and transparency, fostering trust and effective collaboration between domain experts and technical practitioners.

\par In the \textit{Comparison Area} (\autoref{fig:teaser}-\raisebox{-0.13cm}{\includegraphics[height=0.45cm]{Figure/icone2.pdf}}), test set images are displayed vertically, with the first column showing the Grad-CAM generated after the initial training, and each subsequent column presenting the Grad-CAM after each round of fine-tuning. Images with a green border indicate correct label predictions, while a red border indicates incorrect predictions. By observing these changes in local explanations, physicians can more intuitively understand how the focus of the model's attention evolves, gaining deeper insights into the model's decision-making process (\textbf{DG3}).

\section{Evaluation}
\par We evaluated \textit{MEDebiaser} through two studies. First, the \textit{mechanism study} validated that employing human-AI interactive feedback to mitigate existing biases can improve outcomes in the intended direction, providing a quantitative answer to the following research question: \textbf{RQ4:} \textit{Does the system reduce bias in MLMIC and improve model performance, and how do physicians interact with the system to mitigate these biases?}

\par Second, after obtaining institutional IRB approval, we conducted a \textit{user study}, which offered a comprehensive assessment of the user experience with \textit{MEDebiaser}. Our primary goal was to determine whether \textit{MEDebiaser} enhances MLMIC performance by mitigating the effects of \textit{Imbalanced Distribution} and \textit{Label Co-occurrence} and whether it can optimize physician-engineer collaboration, thereby addressing \textbf{RQ4} and the following research question: \textbf{RQ5:} \textit{What’s the impact of the system on workload, physician diagnosis, collaboration, and medical knowledge integration?}

\subsection{Mechanism Study}
\par In this section, we conducted two experiments to evaluate the effectiveness of \textit{MEDebiaser} in mitigating model biases. The first experiment, \textit{With and Without Attention}, compared models that incorporated attention modification on biased images with those that did not. The goal was to determine whether fine-tuning the model using a combination of prediction loss and attention loss effectively mitigated these biases. In the second experiment, \textit{Breadth and Depth}, we explored four different annotation modes, varying in \textit{depth} (few vs. many annotations) and \textit{breadth} (random vs. focused labeling), to analyze how these strategies impact the model's generalization capabilities and its ability to reduce bias.

\subsubsection{Setup and Details}
\par Both experiments were conducted using the \textbf{ChestX-ray14} dataset and carried out on an NVIDIA 4070 Ti GPU. To reduce the annotation workload for physicians, we selected approximately one-tenth of the original dataset while preserving the label distribution and co-occurrence patterns. The data was split into an 8:1:1 ratio for training, validation, and testing. A DenseNet model pretrained on ImageNet was employed for both experiments, with fine-tuning performed using a learning rate of $1e-4$, a batch size of $4$, and $30$ training epochs. To maintain consistency and robustness, we applied standard data augmentation techniques across both experiments.

\subsubsection{Experiment I: With and Without Attention} 
\par \textit{Procedure.} We began by pretraining a DenseNet model on ImageNet, followed by initial training on the \textbf{ChestX-ray14} training set. The model generated chest X-ray images with Grad-CAM heatmaps and corresponding predictions for both the validation and test sets. Physicians were then tasked with reviewing these images, selecting $100$ examples specifically focusing on the label Pleural\_Thickening due to its rarity and higher error rate in prior predictions. Each selected image was meticulously annotated, with polygons drawn around regions of interest, covering all relevant labels.

\par Using these $100$ annotated images, we employed two model fine-tuning techniques. The first method combined prediction loss with attention loss, as outlined in Section~\ref{sec:4.2.2}, leveraging Grad-CAM to refine the model's attention towards the correct regions. The second method relied solely on prediction loss, without any attention-based adjustments.

\begin{table*}[h]
\centering
\caption{Performance comparison for \textit{Experiment I: With and Without Attention.}}
\label{tab:e1}
\begin{tabular}{cccc}
\toprule
\textbf{Metric} & \textbf{Preliminary Training} & \textbf{Prediction Loss Only} & \textbf{Prediction + Attention Loss} \\
\midrule
AUC       & 0.613 & 0.613 & \textbf{0.675} \\
Precision & 0.129 & 0.140 & \textbf{0.160} \\
Recall    & 0.306 & 0.320 & \textbf{0.375} \\
F1 Score  & 0.182 & 0.195 & \textbf{0.224} \\
\bottomrule
\end{tabular}
\end{table*}

\par \textit{Result \& Analysis.} The results, as shown in \autoref{tab:e1}, indicate that the model fine-tuned using the combined prediction loss and attention loss approach demonstrated notable improvements in handling the label Pleural\_Thickening compared to using prediction loss alone. Specifically, the combined approach achieved a \textit{precision} of $0.160$, representing a $14.3\%$ relative increase from $0.140$ under prediction loss only. This improvement is also reflected in \textit{recall} and \textit{F1 score}. In terms of overall performance, \textit{AUC} also showed a meaningful enhancement, improving from $0.613$ with prediction loss only to $0.675$ when incorporating attention loss. This suggests a stronger ability to distinguish between positive and negative cases. These quantitative gains underscore the impact of attention modification, which proved particularly beneficial for underrepresented, tail-end labels. By incorporating attention loss, the model was able to more effectively focus on the relevant image regions associated with such rare labels, reducing false negatives, and addressing the inherent challenges posed by imbalanced multi-label data.

\subsubsection{Experiment II: Breadth and Depth} 
\par \textit{Procedure.} We considered two dimensions for Experiment II: \textit{breadth} and \textit{depth}. \textit{Breadth} refers to annotating a larger number of images, and \textit{depth} focuses on consistently annotating images for a single label. The experimental setup closely mirrored that of Experiment I, utilizing four annotation strategies: 1) annotating $50$ images with Pleural\_Thickening; 2) annotating $100$ images with Pleural\_Thickening; 3) randomly selecting and annotating $50$ images, and 4) randomly selecting and annotating $100$ images. Efforts were made to maintain consistency in the total number of labels covered across these strategies. For each, the model was fine-tuned using a combined approach of prediction and attention loss.

\begin{table*}[h]
\centering
\caption{Performance comparison for \textit{Experiment II: Breadth and Depth.}}
\label{tab:e2}
\begin{tabular}{cccccc}
\toprule
\textbf{Metric} & \textbf{Preliminary Training} & \makecell{\textbf{50 Images} \\ (Focused)} & \makecell{\textbf{100 Images} \\ (Focused)} & \makecell{\textbf{50 Images} \\ (Random)} & \makecell{\textbf{100 Images} \\ (Random)} \\
\midrule
AUC       & 0.613 & 0.620 & \textbf{0.675} & 0.610 & 0.631 \\
Precision & 0.129 & 0.138 & \textbf{0.160} & 0.131 & 0.141 \\
Recall    & 0.306 & 0.312 & \textbf{0.375} & 0.295 & 0.323 \\
F1 Score  & 0.182 & 0.191 & \textbf{0.224} & 0.181 & 0.196 \\
\bottomrule
\end{tabular}
\end{table*}

\par \textit{Result \& Analysis.} As shown in~\autoref{tab:e2}, the model's performance varied depending on the annotation strategy. Strategies involving a higher number of annotated images consistently yielded in higher \textit{precision} and \textit{AUC} compared to those using only $50$ images. Notably, focusing on Pleural\_Thickening, which is underrepresented, led to significant improvements in \textit{recall}. These findings highlight that both annotation \textit{depth} and \textit{breadth} play a crucial role in enhancing the model's generalization capabilities and reducing bias, particularly for tail-end labels.

\subsection{User Study}
\par In this section, we compared the experiences of physicians using \textit{MEDebiaser} with their typical interactions with AI models. To validate the effectiveness of our work in real-world scenarios, we chose to conduct our \textit{user study} in the field of otolaryngology, as endoscopic images of the ear typically display multiple distinct pathological features, and there is a certain degree of correlation between different symptoms.

\subsubsection{Participants}
\par We collaborated with otolaryngology experts from [Blinded for Review], and computer science experts from a local university. We recruited 12 participants, as detailed in \autoref{tab:userparticipants} (physicians: mean age = $40$, SD = $6.56$; engineers: mean age = $32.7$, SD = $4.19$). The group included $6$ otolaryngology experts and $6$ ML experts, with $3$ holding a Master's degree, $5$ holding an M.D., and $4$ holding a Ph.D.. Notably, seven of the participants had prior experience collaborating with counterparts in medical machine learning projects. Three physicians have no background or limited experience with AI. Participants were randomly paired into $6$ teams consisting of one physician and one engineer.

\begin{table*}[htbp]
\begin{minipage}{\textwidth}
  \caption{The details of \textit{User Study} participants.}
  \label{tab:userparticipants}
  \centering
  \begin{tabular}{ccccc|ccccc}
    \toprule
    \multicolumn{5}{c|}{\textbf{Physicians - Otolaryngology}} & \multicolumn{5}{c}{\textbf{Engineers}} \\
    \midrule
    \textbf{ID} & \textbf{Gender} & \textbf{Age} & \textbf{Experience}\footnotemark[1] & \textbf{Degree} & \textbf{ID} & \textbf{Gender} & \textbf{Age} & \textbf{Experience}\footnotemark[2] & \textbf{Degree} \\
    \midrule
    \textbf{UD1} & Male   & 45  & Yes & M.D.   & \textbf{UP1}  & Female & 29  & Yes & Master \\
    \textbf{UD2} & Male   & 38  & No  & M.D.   & \textbf{UP2}  & Male   & 32  & Yes & Ph.D.  \\
    \textbf{UD3} & Female & 50  & Yes & Ph.D.  & \textbf{UP3}  & Female & 27  & Yes & Master \\
    \textbf{UD4} & Male   & 42  & Yes & M.D.   & \textbf{UP4}  & Female & 40  & No  & Ph.D.  \\
    \textbf{UD5} & Female & 35  & No  & M.D.   & \textbf{UP5}  & Male   & 35  & No  & Master \\
    \textbf{UD6} & Male   & 30  & No  & M.D.   & \textbf{UP6}  & Male   & 33  & Yes & Ph.D.  \\
    \bottomrule
      \multicolumn{10}{l}{\footnotesize \textsuperscript{1} Experience stands for Experience with engineers \& AI.} \\
      \multicolumn{10}{l}{\footnotesize \textsuperscript{2} Experience stands for Experience with physicians.} \\
  \end{tabular}

   \end{minipage}
\end{table*}

\subsubsection{Dataset and Model}
\par In the \textit{user study}, we utilized an ear endoscopy dataset provided by participants of the study, sourced from [Blinded for Review]. This dataset was collected from $1,272$ patients between $2018$ and $2022$, comprising a total of $3,252$ images, each with a corresponding diagnostic report. We extracted information from the reports and labeled the associated images. Initially, $12$ labels were present in the dataset, but after excluding labels with insufficient representation, we refined it to include $7$ labels across $3,084$ images, referred to as \textbf{EarEndo}. This dataset exhibits significant issues of \textit{Imbalanced Distribution} and \textit{Label Co-occurrence}, as shown in~\autoref{fig:ear}, Appendix~\ref{sec:User Study} and ~\autoref{tab:earMatrix}, Appendix~\ref{sec:User Study}. In accordance with the hospital's confidentiality policies, patient names, IDs, addresses, and other personal information were anonymized. Due to the limited dataset size and time constraints of the user study, we instructed the physicians to use \textit{DenseNet} with the default parameter settings (batch size = $4$, epoch = $30$) throughout the process. The study was conducted on an NVIDIA 4070 Ti GPU.

\subsubsection{Procedure and Task}

\par
\begin{figure*}[h]
\centering
\includegraphics[width=1\linewidth]{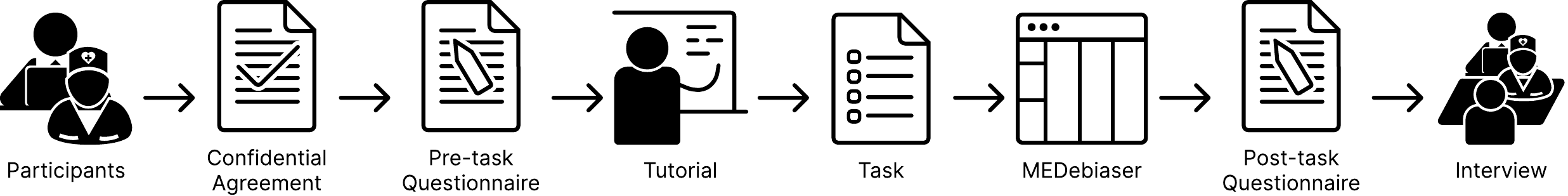}
\caption{The procedure of the \textit{User Study}.}
\label{fig:user}
\end{figure*}

 \autoref{fig:user} illustrates the procedure of our \textit{user study}. Before the study, participants signed privacy and confidentiality agreements and completed a pre-task questionnaire to collect demographic information and task-related background. One author then conducted a 10-minute tutorial on \textit{MEDebiaser} usage and assigned the tasks for the study (\autoref{tab:task}). Then, each team used \textit{MEDebiaser} to mitigate bias related to a specific label. During this period, only physicians operated, while physicians and engineers were restricted from communication except during the \textit{Evaluating Model Performance} stage. The physicians' interactions with \textit{MEDebiaser} were screen recorded, and conversations between physicians and engineers were recorded. Afterwards, participants completed a post-task questionnaire individually. Finally, each team participated in an approximately 20-minute interview. The entire \textit{user study} lasted about one hour on average per participant, who received a \$$10$ token of appreciation upon completion.

\par We carefully chose tasks based on physicians' typical work pattern (\autoref{fig:traditional}) to address key aspects of \textbf{RQ4} and \textbf{RQ5}, in order to measure both technical analysis and collaborative interaction. This approach allowed us to gain a holistic understanding of \textit{MEDebiaser}'s impact on improving both model performance and the user experience.

\begin{table*}
  \caption{Tasks for Using \textit{MEDebiaser}.}
  \label{tab:task}
  \begin{tabular}{ccp{1.6cm}p{9.3cm}}
    \toprule
    \textbf{ID} & \textbf{View} & \textbf{Parties} & \textbf{Tasks} \\
    \midrule
    \textbf{T1} & Label View & Physicians & Describe the distribution and co-occurrence of each label in the dataset. (\textbf{RQ4}) \\
    \midrule
    \textbf{T2} & Attention View & Physicians & Identify biased images, particularly those with infrequently occurring labels and label co-occurrence. (\textbf{RQ4})\\
    \midrule
    \textbf{T3} & Modification View & Physicians & Adjust the attention for images with biased labels. (\textbf{RQ4}, \textbf{RQ5})\\
    \midrule
    \textbf{T4} & Performance View & Physicians \& Engineers & Observe and analyze model results after preliminary training and each round of fine-tuning. (\textbf{RQ4}, \textbf{RQ5})\\
    \bottomrule
  \end{tabular}
\end{table*}
    
\subsubsection{Result Processing and Measurement}
\par In the screen recordings of interactions with \textit{MEDebiaser}, one author was tasked with capturing user activities, including the number and duration of each fine-tuning session. For the audio recordings of conversations between physicians and engineers, as well as participant interviews, we transcribed these into text and conducted a thematic analysis~\cite{10.1093/acprof:oso/9780199753697.001.0001} by coding the transcripts. The time spent on model training and fine-tuning was not included.

\par For the post-task questionnaire, a 7-point Likert scale was employed ($1$: Not at all/Strongly disagree, $7$: Very much/Strongly agree). Further details are available in \autoref{tab:survey_questions}. The questions were categorized into four aspects: \textit{bias detection}, \textit{system usability scale}~\cite{10.5555/2817912.2817913}, \textit{feedback \& communication}, and \textit{workload}~\cite{HART1988139}. Among these, the \textit{system usability scale} and \textit{bias detection} primarily addressed \textbf{RQ4}, while \textit{feedback \& communication} and \textit{workload} primarily addressed \textbf{RQ5}. Even though engineers do not directly interact with the system, their feedback on bias detection, usability, and workload can provide valuable insights into the system's overall performance in a collaborative setting. Thus, as observers, they also filled out the same questionnaire.

\par For \textbf{T1}, the authors rated each participant's description of the dataset on a 5-point Likert scale based on its alignment with the actual data. For \textbf{T2} and \textbf{T3}, we analyzed the recorded interactions, counting the number of tasks completed, as well as the time and speed taken. For \textbf{T4}, we combined the scores from relevant questions in the post-task questionnaire with the analysis of the audio recordings to assess participants' completion of the task and filter valid questionnaires.

\begin{table*}
  \caption{Details of the post-task questionnaire.}
  \label{tab:survey_questions}
  \begin{tabular}{p{4.5cm}p{3.5cm}p{8.5cm}}
    \toprule
    \textbf{Aspect} & \textbf{Topic} & \textbf{Question} \\
    \midrule
    Bias Detection (\textbf{RQ4})
    & Imbalance Mitigation & Does \textit{MEDebiaser} show improvement in mitigating imbalanced label distributions compared to using the model directly? \\
    & Co-occurrence Distinction & Does \textit{MEDebiaser} show improvement in distinguishing co-occurring labels compared to using the model directly? \\
    \midrule
    System Usability Scale (\textbf{RQ4})
    & Ease of Learn & How easy was it to learn to use \textit{MEDebiaser}? \\
    & Ease of Use & How user-friendly do you find \textit{MEDebiaser}'s interface and interactions? \\
    & Likelihood of Future Use & How likely are you to recommend using \textit{MEDebiaser} in the future? \\
    \midrule
    Feedback \& Communication (\textbf{RQ5})
    & AI to Physicians & How clear and intuitive is the feedback provided by the AI model to physicians? \\
    & Physicians to AI & How effectively can physicians provide human knowledge to improve the AI model? \\
    & Between Physicians and Engineers & How well does \textit{MEDebiaser} facilitate collaboration between physicians and engineers? \\
    \midrule
    Workload (\textbf{RQ5}) 
    & Time Aspect & Does using \textit{MEDebiaser} reduce the time it takes to complete tasks? \\
    & Physical Aspect & Does using \textit{MEDebiaser} reduce the physical demands? \\
    \bottomrule
  \end{tabular}
\end{table*}

\subsection{User Study Results}

\par We compiled information on users' interactions with \textit{MEDebiaser} through the study. Initially, all physicians successfully uploaded their datasets and selected a model in the \textit{Panel View}. They then proceeded to the \textit{Label View}, where five physicians sorted the table in ascending order by distribution percentage. These physicians selected categories with lower distribution, such as EOM, OME, and EOF, and closely examined the label co-occurrence in the chord diagram. In the \textit{Attention View}, they reviewed $38$ images with a total of $49$ labels under EOF. Among these images, $18$ were correctly predicted, covering $25$ labels, while $20$ images were incorrectly predicted. Four physicians zoomed in on each image to correct biases in unreasonable predictions, with \textbf{UD3} and \textbf{UD4} opting to directly select correctly predicted attention and adjust only for incorrect predictions. In the \textit{Modification View}, all six physicians adeptly used polygons to annotate the images, and \textbf{UD1} and \textbf{UD5} also utilized the dropdown menu to annotate the remaining labels. For recommendation, \textbf{UD1}, \textbf{UD2}, and \textbf{UD6} chose the dependency mode for sorting, \textbf{UD3} and \textbf{UD5} selected concentration, and \textbf{UD4} opted for accuracy. In the \textit{Performance View}, the physicians reviewed EOF, carefully comparing each image with the results from the previous round. During this stage, the physicians discussed the model's metrics and collaborated with engineers to get further assurance.

\subsubsection{RQ4: Does the system reduce bias in MLMIC and improve model performance, and how do physicians interact with the system to mitigate these biases?}

\par \textbf{Using \textit{MEDebiaser}, users gain a better understanding of their datasets, can more easily identify biased images, make necessary adjustments, and ultimately enhance the model's performance.} During the interview, physicians effectively described the overall label distribution in \textbf{EarEndo} and the co-occurrence of specific labels (\textbf{T1}), with all participants receiving good scores in the rating, offering insights from a medical perspective. Through the \textit{Attention View}, which presents local explanations, physicians reported that they could swiftly identify images with incorrect predictions (\textbf{T2}). \textbf{UD2} noted, ``\textit{The explanations for correctly predicted images usually make sense, but sometimes they aren't detailed [enough] and might need some tweaks. For images that were predicted incorrectly, the explanations are obviously off, so I focus on fixing those [first].}''

\par
\begin{figure}[h]
\centering
\includegraphics[width=1\linewidth]{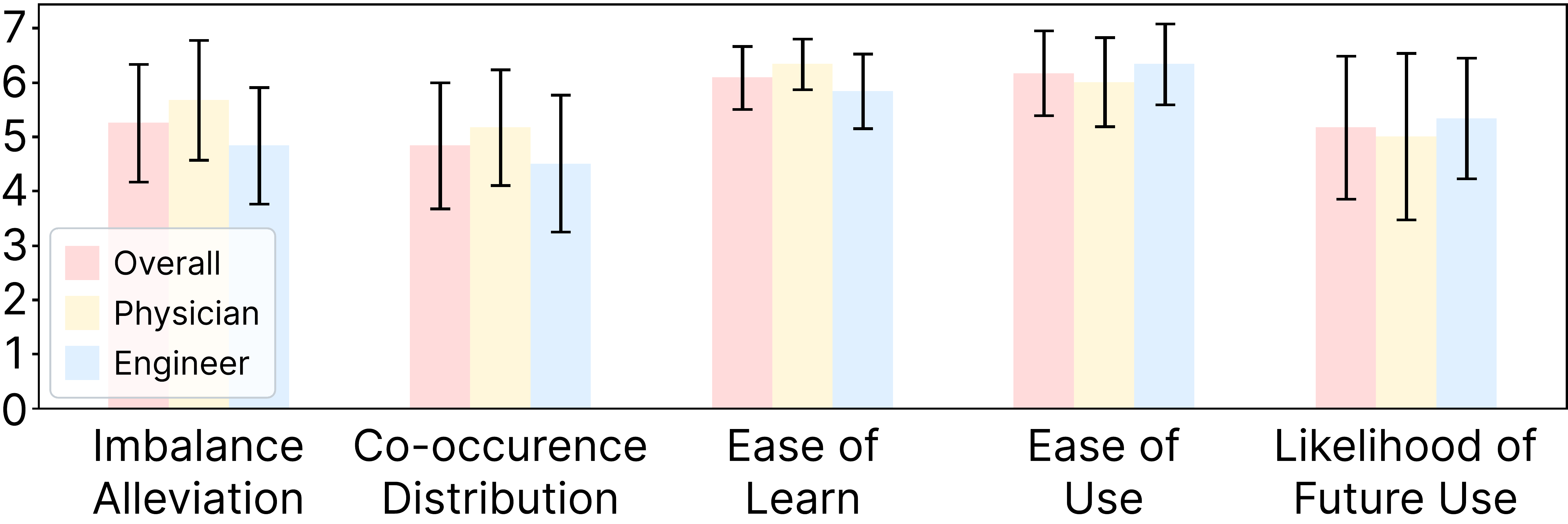}
\caption{Likert Results on \textit{Bias Detection} and \textit{System Usability Scale}.}
\label{fig:q1_5}
\end{figure}

 Regarding imbalance mitigation (physicians: mean = $5.67$; engineers: mean = $4.83$) and co-occurrence distinction (physicians: mean = $5.17$; engineers: mean = $4.5$), it is clear from \autoref{fig:q1_5} that physicians rated these aspects higher than engineers. Several physicians observed that, in the \textit{Performance View}, after multiple rounds of fine-tuning, the attention seems to align more accurately with the correct regions. However, two engineers mentioned that the relatively small amount of annotated data in the \textit{user study} has led to only minor changes in metrics. \textbf{UP1} commented, ``\textit{Right now, with the limited number of annotations, the metrics only show small improvements. But I think as we add more [annotations], we'll start to see a bigger reduction [in bias].}''

\par
\begin{figure}[h]
\centering
\includegraphics[width=1\linewidth]{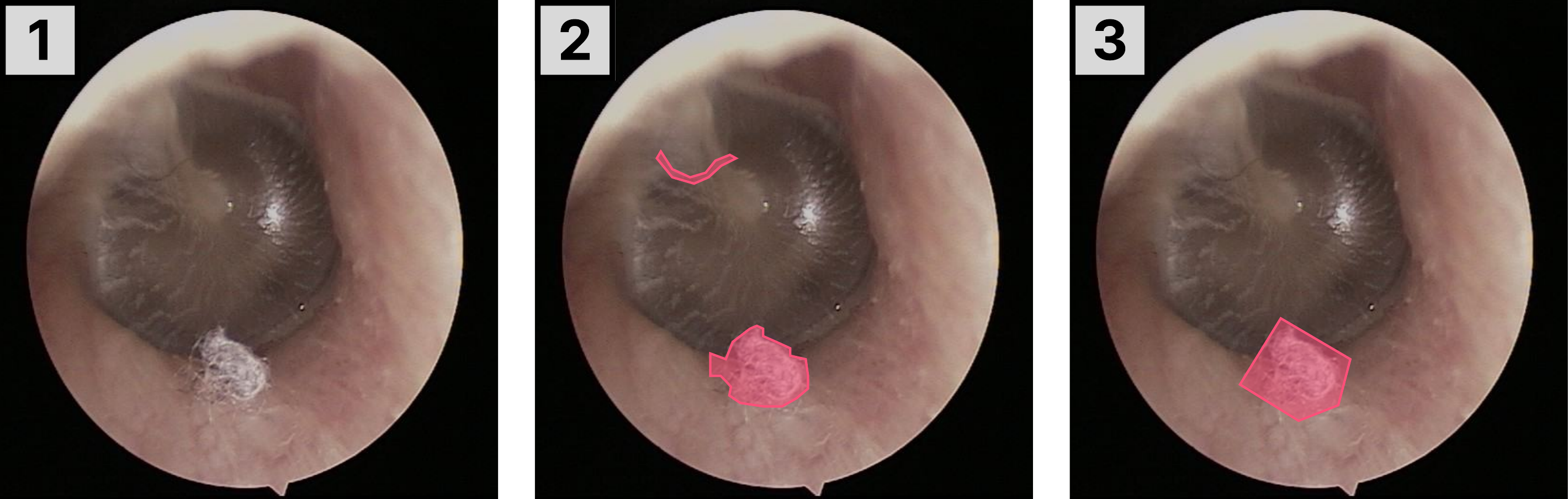}
  \caption{\raisebox{-0.07cm}{\includegraphics[height=0.35cm]{Figure/icon1.pdf}} Original Image.  \raisebox{-0.07cm}{\includegraphics[height=0.35cm]{Figure/icon2.pdf}} Polygon annotated by \textbf{UD5}. \raisebox{-0.07cm}{\includegraphics[height=0.35cm]{Figure/icon3.pdf}} Polygon annotated by \textbf{UD4}.}
  \label{fig:userear}
\end{figure}

\begin{figure}[t]
\centering
\includegraphics[width=1\linewidth]{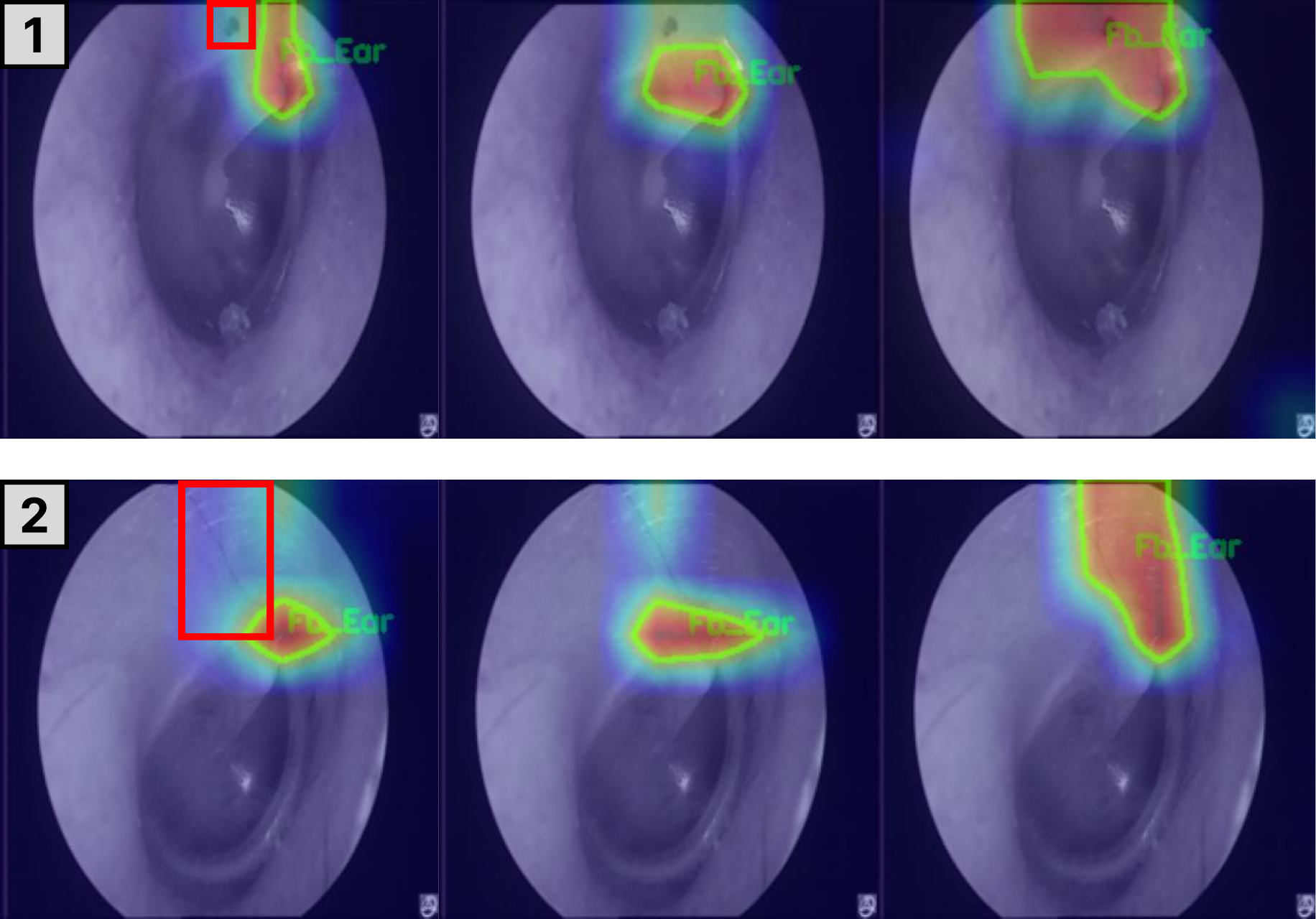}
\caption{The results of two rounds of fine-tuning on EOF by \textbf{UD1}. In \raisebox{-0.07cm}{\includegraphics[height=0.35cm]{Figure/icon1.pdf}}, the red box contains a black foreign object. After each fine-tuning, the attention gets closer to the foreign object. After two rounds of fine-tuning, the attention covers the foreign object. In \raisebox{-0.07cm}{\includegraphics[height=0.35cm]{Figure/icon2.pdf}}, the red box contains a strand of hair. After each fine-tuning, the attention gets closer to the hair. After two rounds of fine-tuning, the attention covers the entire hair.}
\label{fig:example}
\end{figure}

 \textbf{\textit{MEDebiaser} is easy to use, and its interactive human-AI feedback mechanism effectively aids users in correcting model biases with ease.} Both physicians and engineers acknowledged the ease of use of \textit{MEDebiaser} (physicians: mean = $6$; engineers: mean = $6.33$) and its low learning curve (physicians: mean = $6.33$; engineers: mean = $5.83$), as shown in \autoref{fig:q1_5}. Notably, \textbf{UD2}, \textbf{UD5}, and \textbf{UD6}, despite having no prior ML experience, also found \textit{MEDebiaser} accessible and agreed that it requires minimal ML knowledge to operate. \textbf{UD3} mentioned, ``\textit{MEDebiaser is quite user-friendly and easy to use. It's similar to Labelme~\cite{Russell2008LabelMeAD}, which I have used before for image annotation.}'' As shown in~\autoref{tab:finetune}, all six teams completed the modification work for EOF within the designated time (\textbf{T3}). Among them, \textbf{UD4} completed the task with only three rounds of fine-tuning in the shortest time, while \textbf{UD5} took the longest, finishing in five rounds. This difference is attributed to \textbf{UD4}'s preference for creating rough outlines with fewer points, compared to \textbf{UD5}'s more detailed approach with precise points, as shown in~\autoref{fig:userear}. This iterative feedback mechanism allowed physicians to observe changes and trends in the model, facilitating dynamic adjustments and enhancing efficiency. As \textbf{UD1} noted, ``\textit{After two rounds of adjustments, I could see the results getting better (\autoref{fig:example}). I was thinking of moving on to other [labels], but I ended up sticking with the remaining [EOF] images as needed.}'' Overall, both physicians and engineers expressed a willingness to use \textit{MEDebiaser} in the future (physicians: mean = $5$; engineers: mean = $5.33$).

\begin{table*}
\begin{minipage}{\textwidth}
  \caption{The details of using \textit{MEDebiaser} on EOF.}
  \label{tab:finetune}
  \begin{tabular}{ccccccccccccc}
    \toprule
     & \multicolumn{2}{c}{\textbf{Round 1}} & \multicolumn{2}{c}{\textbf{Round 2}} & \multicolumn{2}{c}{\textbf{Round 3}} & \multicolumn{2}{c}{\textbf{Round 4}} & \multicolumn{2}{c}{\textbf{Round 5}} & \multicolumn{2}{c}{\textbf{Total}} \\
    \cmidrule(r){2-3} \cmidrule(r){4-5} \cmidrule(r){6-7} \cmidrule(r){8-9} \cmidrule(r){10-11} \cmidrule(r){12-13}
    & \textbf{Num.} & \textbf{Time} & \textbf{Num.} & \textbf{Time} & \textbf{Num.} & \textbf{Time} & \textbf{Num.} & \textbf{Time} & \textbf{Num.} & \textbf{Time} & \textbf{Num.} & \textbf{Time} \\
    \midrule
    \textbf{UD1} & 8 & 177s & 10 & 184s & 13 & 300s & 9 & 165s & 9 & 200s & 49 & 1,026s \\
    \textbf{UD2} & 7 & 149s & 5 & 108s & 6 & 124s & 7 & 137s & --- & --- & 25 & 518s \\
    \textbf{UD3} & 6 & 138s & 4 & 85s & 5 & 129s & 3 & 67s & --- & --- & 18 & 419s \\
    \textbf{UD4} & 5 & 102s & 7 & 155s & 6 & 131s & --- & --- & --- & --- & 18 & 388s \\
    \textbf{UD5} & 12 & 278s & 9 & 213s & 13 & 305s & 8 & 172s & 7 & 159s & 49 & 1,127s \\
    \textbf{UD6} & 4 & 84s & 6 & 137s & 8 & 188s & 3 & 65s & --- & --- & 23 & 474s \\
    \bottomrule
      \multicolumn{13}{l}{\footnotesize \textbf{Num.} stands for the number of labels.} \\
  \end{tabular}
     \end{minipage}
\end{table*}

\subsubsection{RQ5: What’s the impact of the system on workload, physician diagnosis, collaboration, and medical knowledge integration?}

\par
\begin{figure}[h]
\centering
\includegraphics[width=1\linewidth]{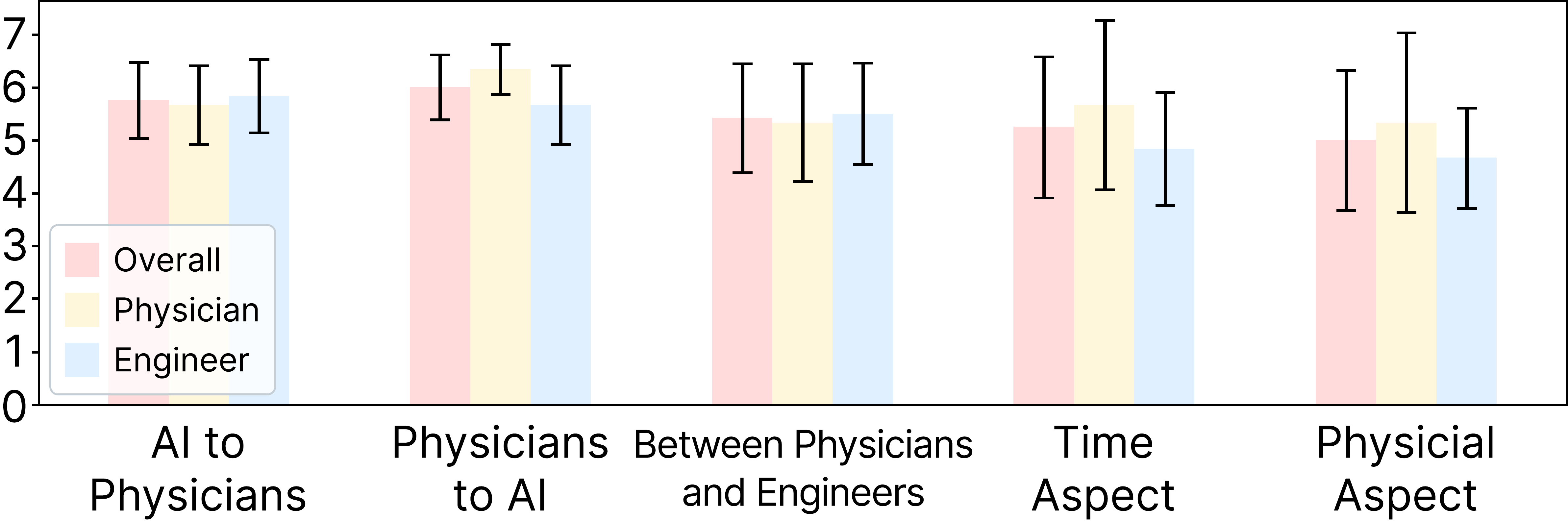}
\caption{Likert Results on \textit{Feedback \& Communication} and \textit{Workload}.}
\label{fig:q5_10}
\end{figure}

 \textbf{\textit{MEDebiaser}'s interface effectively facilitates iterative feedback between physicians and AI models, promoting the integration of expertise into the training process.} \textit{MEDebiaser} employs Grad-CAM to provide local explanations. Despite being a conventional method, both physicians and engineers found this visualization straightforward and easy to interpret (physicians: mean = $5.67$; engineers: mean = $5.83$), as shown in \autoref{fig:q5_10}. \textbf{UD6} pointed out, ``\textit{We frequently request visualizations, but sometimes they end up being overly complicated and hard to use. Most of the time, simple heatmaps do the job perfectly well.}'' Regarding the integration of medical expertise, physicians expressed high satisfaction (mean = $6.33$). \textbf{UD1} mentioned, ``\textit{Previously, it was really challenging for me to communicate the important [medical] features I had in mind to engineers and have them implemented. However, with this new approach, my ideas are now directly and accurately represented in the model, and the heatmaps and line charts it produces are clear and easy to understand.}'' (\textbf{T2}, \textbf{T4}). Engineers found this method of knowledge integration more accessible to physicians and less complex compared to existing approaches (mean = $5.67$). \textbf{UP3} commented, ``\textit{MEDebiaser's interface is designed with clarity, making it easy to interpret the model and guide users effectively. We often use iterative training methods in our work, but usually lack a user interface to facilitate these [processes].}''

\par \textbf{With \textit{MEDebiaser}, the physician-engineer collaboration has shifted from the traditional, lengthy ``Bias$\leftrightarrows$Revision'' cycle to a more dynamic mode where physicians take the lead, with engineers offering support.} Both engineers and physicians generally agree that \textit{MEDebiaser} facilitates their collaboration (physicians: mean = $5.33$; engineers: mean = $5.5$). \textbf{UP5} mentioned, ``\textit{This essentially lets doctors make initial adjustments before they bring up any issues to us. If it delivers the results we're aiming for, it definitely makes our job easier.}'' \textbf{UP2} added, ``\textit{This approach helps break [down] some of the communication barriers between us and them. Sometimes we don't completely grasp the specific outcomes they're looking for. Now, they can make initial corrections on their own, without having to depend on us for everything.}'' In the traditional mode, physicians often served as AI users, while engineers acted as the AI providers, creating a demand-driven ``client-provider'' dynamic where engineers held most of the responsibility for managing and maintaining the models. \textit{MEDebiaser} shifts this paradigm, fostering joint management and maintenance of AI by physicians and engineers as equal collaborators. With this tool, engineers are no longer tasked solely with ongoing modifications and adjustments---many of these responsibilities are now shared with physicians in a way that aligns with their expertise. Physicians, in turn, are empowered to apply their domain knowledge more effectively, without being constrained by the knowledge gap between the two fields. This collaborative mode complements, rather than replaces, the traditional ``Bias$\leftrightarrows$Revision'' cycle. Physicians can still seek assistance from engineers when needed, but this new framework allows both parties to utilize their strengths more efficiently and productively.

\par \textbf{\textit{MEDebiaser} has effectively reduced both the time and physical workload for physicians and engineers.} Most physicians found its interactive mechanism significantly decreased the number of required annotations compared to annotating the entire dataset before training, reducing time (mean = $5.67$) and physical workload (mean = $5.33$). However, \textbf{UD2} commented, ``\textit{Even though it cuts down on the number of annotations compared to labeling the entire dataset, I still don't want to spend too much [time] on this annotation [work].}'' Regarding the physical workload, physicians had varying opinions. \textbf{UD4}, who used fewer points for each annotation, felt the physical workload was manageable, while \textbf{UD5}, who meticulously annotated each image, expressed concern: ``\textit{When there's a lot of data, this work can get exhausting, both physically and mentally.}'' engineers also reported reduced time workload (mean = $4.83$) and physical workload (mean = $4.67$). \textbf{UP1} noted, ``\textit{Our main role is to guide the training settings and confirm the results, which is a lot easier than having to fix the model whenever something goes wrong, especially when we don't fully understand the data. }''

\section{Discussion}

\subsection{Design Implications}
\subsubsection{Division of Responsibilities Between Physicians and Engineers}
\par In complex medical human-AI collaboration, clearly defining the roles of physicians and engineers is crucial for effective outcomes. In \textit{MEDebiaser}, physicians go beyond annotating data by interpreting model outputs and understanding how their feedback influences training, as noted by \textbf{UD2}: ``\textit{It's important for me to not only provide feedback but also understand how it impacts the model's behavior over time.}'' Meanwhile, engineers handle the technical implementation, ensuring model performance and explaining its behavior to physicians. They maintain control over the model’s integrity and adjust algorithms as needed.
Regardless of the specific collaboration mode, the key principle is aligning roles with each party’s expertise, ensuring continuous feedback, and reducing dependency. Such collaboration leads to more effective human-AI solutions. Future systems should integrate this to maximize both human expertise and AI capabilities for improved healthcare outcomes.

\subsubsection{Human Workload, AI Performance, and Their Tradeoff}
\par In \textit{MEDebiaser} and similar AI systems, the human workload is a critical factor in determining the system's scalability. Unlike traditional annotation methods that require labeling the entire dataset, \textit{MEDebiaser} reduces this burden by focusing specifically on biased images and offering customized recommendation strategy to further ease the workload. In the \textit{user study}, \textbf{UD5} raised concerns about the increasing manual annotation efforts required for improving model accuracy and mitigating bias. These concerns are valid, as manual annotation can be taxing for physicians, especially with large datasets. However, our work targets the underrepresented tail-end labels that contribute to model bias, aiming to enhance model fairness and performance without overwhelming users with excessive manual labor.

\par Another challenge in the medical domain, unlike natural image classification, is the scarcity of data for rare classes, where features are often sparse and difficult to capture, even with state-of-the-art models. While no tool can fully compensate for the lack of data, the most effective solution is to increase the number of cases. \textit{MEDebiaser} works within the real-world data constraints, leveraging physician involvement to help the model focus on rare classes. The data distribution used in our experiments aligns with that in most research on imbalanced datasets, and both the \textit{mechanism study} and \textit{user study} validate the effectiveness of our approach. However, it is important to acknowledge that the limitations of sparse medical data will still impact model performance, and our approach is an effective attempt to partially mitigate this issue.

\par In the \textit{MEDebiaser} framework, while annotating additional data enhances model performance, it may also increase the workload for physicians. To mitigate this, our system employs a customized ranking strategy that prioritizes the most critical and underrepresented labels, thereby reducing the amount of annotation required. This approach helps minimize manual effort while improving both model accuracy and fairness. However, there is potential for further optimization by incorporating more advanced automation techniques, such as diffusion learning~\cite{diagnostics14131442}. Diffusion learning can facilitate the transfer of patterns from annotated to unannotated images, enabling the model to automatically infer labels and reduce the need for manual annotations~\cite{diagnostics14131442}. Physicians could then review and refine these auto-generated annotations, further alleviating their workload. Similar strategies have been successfully applied in medical text classification tools~\cite{10540286,LI2021345}, where expert knowledge and machine learning work collaboratively to improve efficiency. However, the use of such technologies requires significant design and experimentation to ensure their effectiveness and safety. Therefore, we have not explored or attempted this approach further in our work.

\subsubsection{Human Autonomy, AI Controllability, and Their Tradeoff}
\par In the \textit{user study}, we observed that some participants, such as \textbf{UD4}, exhibited behaviors indicative of a group that prefers to delegate much of the task to AI, especially for tasks like image segmentation and detection, which they believe are already highly mature. \textbf{UD4} completed tasks quickly, with fewer annotations and polygon vertices. \revise{This behavior, however, represents one end of a spectrum of interaction styles we observed. In stark contrast, other physicians adopted a more meticulous and cautious approach, creating highly detailed annotations with numerous vertices, reflecting a desire to maintain greater control over the process. This potential variability in physician annotation behavior, stemming from differing levels of trust in AI and personal diligence, is a critical factor to consider. The sentiment from \textbf{UD4}, who mentioned during the interview, ``\textit{AI is already powerful enough, just let it handle the task,}'' exemplifies the ``delegator'' attitude and reflects a common inclination in medical settings to over-rely on AI for tasks considered well-suited to automation.} However, while the potential for automation in healthcare is evident, its feasibility largely depends on the accuracy of the model, as a fully automated system cannot be guaranteed to achieve 100\% accuracy.

\par While AI systems have demonstrated remarkable performance in tasks like medical image segmentation and classification, it is crucial to recognize that these models' effectiveness is often dependent on the data distribution used during training. Consequently, these models may struggle when confronted with rare or atypical cases. This highlights the continuing need for human involvement. Physicians should not be expected to compete with AI in tasks where it excels, such as image segmentation or classification. However, their expertise is indispensable in areas where AI may falter---such as interpreting ambiguous findings, incorporating patient history, and identifying subtle cues beyond the model's scope. In these instances, human oversight ensures AI's performance aligns with real-world clinical requirements, particularly in handling rare or minority cases that may be misclassified due to data imbalance. Therefore, even in highly automated settings, human intervention remains necessary to validate and guide AI-driven decisions.


\par Currently, \textit{MEDebiaser} offers visual feedback on intermediate results, providing some transparency into the model's behavior. However, this doesn't fully ensure understanding or control, as physicians may struggle to interpret the data and determine the right parameter adjustments, potentially disrupting the training process. To address this, future versions could include an automatic recommendation system that suggests optimal parameter settings based on training progress, as well as anomaly detection and real-time alerts to notify physicians of irregularities. These enhancements would enable physicians to monitor and control the model more effectively, ensuring that training remains on track and that optimal adjustments are made.



\subsection{Generalizability}
\subsubsection{Data Update}
\par In the formative study, \textbf{D3} remarked: ``\textit{We see a large number of patients every day, generating new images and diagnoses. It would be ideal if this data could be utilized.}'' The continuous influx of new data presents a challenge in keeping the model up-to-date and adaptive to evolving trends. \textit{MEDebiaser} addresses this by recommending the most valuable data points for physicians to annotate, particularly when new data becomes available. By identifying uncertain or unfamiliar cases, the system minimizes effort spent on well-understood samples and instead prioritizes those that contribute most to model improvement. This targeted recommendation streamlines the annotation process while ensuring the model remains aligned with the latest medical knowledge, enhancing its generalizability over time. Consequently, \textit{MEDebiaser} is well-suited for deployment in hospital environments where data is continuously updated.


\subsubsection{\revise{Broader Applications in Medical and Non-Medical Scenarios}}
\par In real-world scenarios, multi-label data is common. \textit{MEDebiaser} has shown effectiveness in reducing bias in such datasets, making it suitable for broader applications across various complex, multi-label contexts. What sets \textit{MEDebiaser} apart from other state-of-the-art MLMIC models is its interactive design, allowing integration of human expertise during training, unlike models where physicians have to passively accept outputs. This interactivity boosts its utility in medical contexts and beyond. \revise{While our evaluation focuses on two datasets, \textit{MEDebiaser} is applicable to other medical multi-label tasks because its core workflow is designed to address fundamental challenges, like imbalanced distributions and label co-occurrence, that are common across many medical imaging domains, such as the analysis of skin lesions in dermatology, tissue classification in digital pathology, and the identification of multiple findings in brain MRI scans.} Additionally, while medical annotations depend on domain experts, natural image datasets typically do not require such specialized expertise for annotations. In these cases, crowdsourcing~\cite{10.1117/1.JMI.5.3.034002} can harness public input, and when combined with AI systems like \textit{MEDebiaser}, generate accurate and reliable labels even without specialized knowledge.

\subsection{Limitation \& Future Work}

\subsubsection{Recommendation Strategy}
\par The current recommendation strategy in \textit{MEDebiaser} uses a ranking method based on label accuracy, heatmap concentration, and co-occurrence matrix dependency, focusing on model-driven metrics over medical ones. \textbf{UD3} noted, ``\textit{The system's recommendations are helpful, but they don't always match what we prioritize in clinical practice. It would be better if it could suggest cases based on medical similarities rather than just relying on model outputs.}'' Physicians prefer recommendations based on medical feature similarity to better analyze specific symptoms. Future work could develop a dynamic recommendation system grounded in medical features and tailored to physicians' annotation patterns, helping them identify model issues more effectively. Additionally, recommendations could be tailored to physicians' actual annotation patterns. For example, if a certain region is frequently annotated, it may suggest that symptoms in that area are more challenging to recognize or prone to model misclassification. In such cases, the system could prioritize recommending images where the Grad-CAM heatmap does not highlight that region, thereby helping physicians more efficiently identify potential model issues and areas for improvement.

\subsubsection{\revise{Practical Deployment Considerations}}
\par \revise{Moving \textit{MEDebiaser} from a research prototype to a clinical tool involves several practical considerations. For data privacy, the system is designed for on-premises deployment within a hospital's secure network to comply with strict regulations. Regarding hardware, while the front-end is a lightweight web application, the back-end requires a server with a modern GPU for efficient model fine-tuning. Physician training must be brief and practical, and the tool must be integrated into clinical workflows as a retrospective activity requiring dedicated time. Beyond these, a clear protocol for model maintenance and versioning is crucial; engineers must validate each new physician-tuned model against a hold-out dataset before it is promoted for wider use. Finally, for long-term adoption, future work should address interoperability with existing hospital systems, such as integrating with Picture Archiving and Communication System (PACS) to streamline image selection and review, further embedding the tool into the natural clinical workflow.}

\section{Conclusion}
\par This study presents \textit{MEDebiaser}, an interactive system designed to mitigate bias in MLMIC. \textit{MEDebiaser} enhances human-AI collaboration by facilitating continuous feedback between physicians and AI models, offering interpretability, and enabling physicians to directly adjust the model's attention during iterative fine-tuning through an intuitive interface. At the same time, it ensures effective co-supervision of AI models between physicians and engineers. By streamlining the physician-engineer collaboration process, \textit{MEDebiaser} reduces the workload for both parties. Our \textit{mechanism study} and \textit{user study} demonstrate that \textit{MEDebiaser} significantly reduces bias in MLMIC, improves usability, and boosts the overall efficiency of physician-engineer collaboration.


\begin{acks}
\par We thank anonymous reviewers for their valuable feedback. This work is supported by grants from the National Natural Science Foundation of China (No. 62372298), the Shanghai Siming Medical Development Foundation (No. SGKY-202405), Shanghai Engineering Research Center of Intelligent Vision and Imaging, Shanghai Frontiers Science Center of Human-centered Artificial Intelligence (ShangHAI), and MoE Key Laboratory of Intelligent Perception and Human-Machine Collaboration (KLIP-HuMaCo).
\end{acks}

\balance
\bibliographystyle{ACM-Reference-Format}
\bibliography{sample-manuscript}


\appendix
\newpage
\section{Content Analysis Supplementary}
\label{sec:content}
See~\autoref{tab:annotation_methods} and~\autoref{tab:usability_results}.

\begin{table*}[h]
\centering
\caption{Summary of Common Annotation Methods in Medical AI Tools}
\begin{tabular}{>{\raggedright\arraybackslash}m{3cm} >{\raggedright\arraybackslash}m{4.2cm} >{\centering\arraybackslash}m{2cm} >{\raggedright\arraybackslash}m{3cm} >{\raggedright\arraybackslash}m{3cm}}
\toprule
\textbf{Annotation Method} & \textbf{Description} & \textbf{Example} & \textbf{Advantages} & \textbf{Disadvantages} \\
\midrule
Bounding Boxes     & A rectangular box drawn around the object.    & \includegraphics[width=1.5cm]{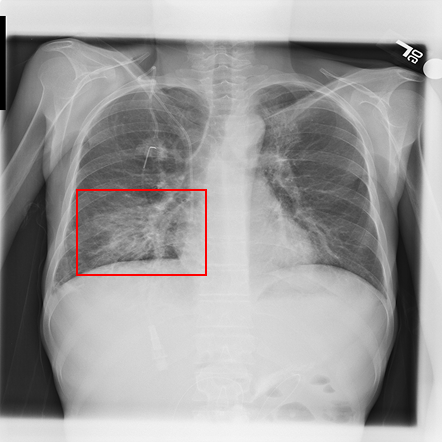}    & Simple to implement; widely supported & Inaccurate for irregular anatomical regions \\
Brush / Flood Fill & Filling a region with color to define an area.    & \includegraphics[width=1.5cm]{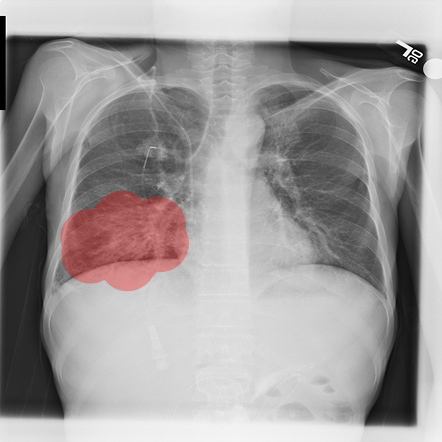}    & Intuitive interaction; pixel-level detail & May produce inconsistent results \\
Polygon            & A multi-point closed shape drawn around an object.    & \includegraphics[width=1.5cm]{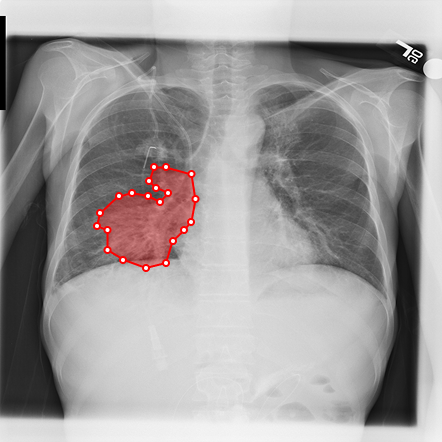}    & High precision for complex contours   & Time-consuming for manual annotation \\
Keypoint Skeleton  & Marking specific key points and connecting them with lines.    & \includegraphics[width=1.5cm]{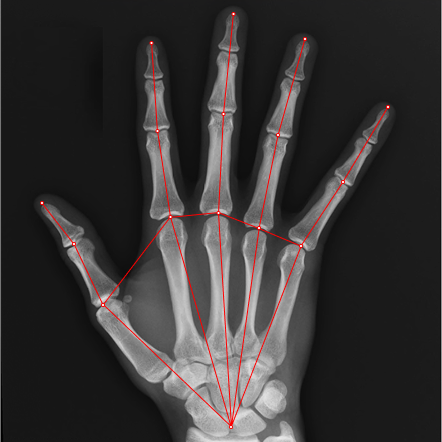}    & Suitable for landmark/pose annotation & Not applicable to region segmentation \\
Polyline           & A series of connected straight lines defining an object.    & \includegraphics[width=1.5cm]{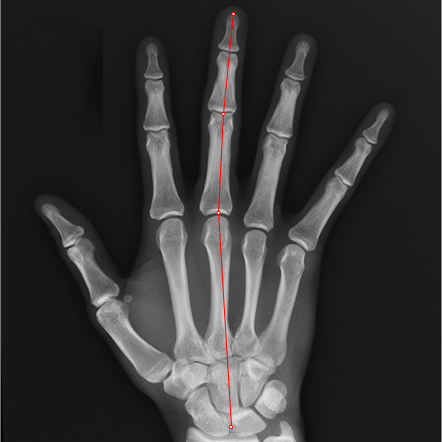}    & Useful for vessel/trachea tracing     & Limited to linear structures \\
\bottomrule
\end{tabular}
\label{tab:annotation_methods}
\end{table*}

\begin{table*}[h]
\centering
\caption{Results of Usability Evaluation on Annotation Methods}
\begin{tabular}{ccccc}
\toprule
\textbf{Annotation Method} & \textbf{Avg. Time (s)} & \textbf{Accuracy (0–1)} & \textbf{Satisfaction (1–5)}\\
\midrule
Bounding Boxes     & \textbf{12.4} & 0.78 & 3.2\\
Polyline  & 15.6 & 0.71 & 3.0\\
Polygon            & 19.7 & \textbf{0.92} & \textbf{4.5}\\
Brush / Flood Fill & 20.5 & 0.89 & 4.0\\
\bottomrule
\end{tabular}
\label{tab:usability_results}
\end{table*}

\section{Loading Dataset and Model Supplementary}
\label{sec:Dataset}
\begin{figure*}[h]
  \centering
  \includegraphics[width=0.6\textwidth]
  {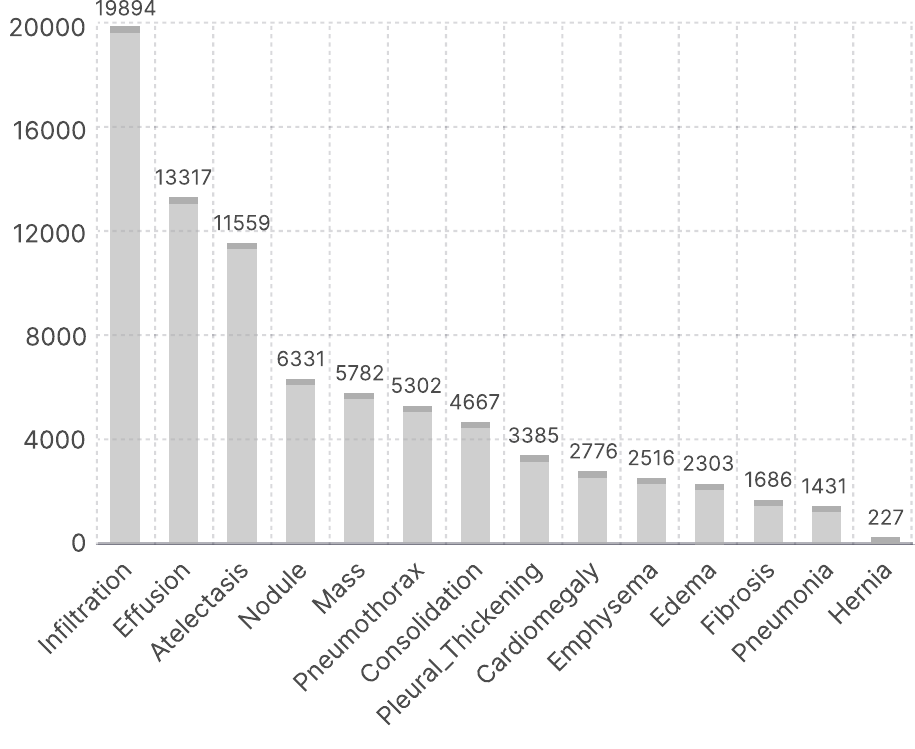}
  \caption{The label distribution of \textbf{ChestX-ray14}.}
  \label{fig:distribution}
\end{figure*}

See~\autoref{fig:distribution} and~\autoref{tab:Matrix}.

\begin{table*}[h]
\begin{minipage}{\textwidth}
\centering
\caption{The co-occurrence matrix of \textbf{ChestX-ray14}.}
\label{tab:Matrix}
\begin{tabular}{lcccccccccccccc}
\toprule
& \textbf{Atl} & \textbf{Car} & \textbf{Con} & \textbf{Ede} & \textbf{Eff} & \textbf{Emp} & \textbf{Fib} & \textbf{Her} & \textbf{Inf} & \textbf{Mas} & \textbf{Nod} & \textbf{Ple} & \textbf{Pne} & \textbf{Ptx} \\
\midrule
\textbf{Atl} & --- & 370 & 1,223 & 221 & 3,275 & 424 & 220 & 40 & 3,264 & 739 & 590 & 496 & 262 & 774 \\
\textbf{Car} & 370 & --- & 169 & 127 & 1,063 & 44 & 52 & 7 & 587 & 102 & 108 & 111 & 41 & 49 \\
\textbf{Con} & 1,223 & 169 & --- & 162 & 1,287 & 103 & 79 & 4 & 1,221 & 610 & 428 & 251 & 123 & 223 \\
\textbf{Ede} & 221 & 127 & 162 & --- & 593 & 30 & 9 & 3 & 981 & 129 & 131 & 64 & 340 & 33 \\
\textbf{Eff} & 3,275 & 1,063 & 1,287 & 593 & --- & 359 & 188 & 21 & 4,000 & 1,254 & 912 & 849 & 269 & 996 \\
\textbf{Emp} & 424 & 44 & 103 & 30 & 359 & --- & 36 & 4 & 449 & 215 & 115 & 151 & 23 & 747 \\
\textbf{Fib} & 220 & 52 & 79 & 9 & 188 & 36 & --- & 8 & 345 & 117 & 166 & 176 & 11 & 80 \\
\textbf{Her} & 40 & 7 & 4 & 3 & 21 & 4 & 8 & --- & 33 & 25 & 10 & 8 & 3 & 9 \\
\textbf{Inf} & 3,264 & 587 & 1,221 & 981 & 4,000 & 449 & 345 & 33 & --- & 1,159 & 1,546 & 750 & 605 & 946 \\
\textbf{Mas} & 739 & 102 & 610 & 129 & 1,254 & 215 & 117 & 25 & 1,159 & --- & 906 & 452 & 71 & 431 \\
\textbf{Nod} & 590 & 108 & 428 & 131 & 912 & 115 & 166 & 10 & 1,546 & 906 & --- & 411 & 70 & 341 \\
\textbf{Ple} & 496 & 111 & 251 & 64 & 849 & 151 & 176 & 8 & 750 & 452 & 411 & --- & 48 & 289 \\
\textbf{Pne} & 262 & 41 & 123 & 340 & 269 & 23 & 11 & 3 & 605 & 71 & 70 & 48 & --- & 41 \\
\textbf{Ptx} & 774 & 49 & 223 & 33 & 996 & 747 & 80 & 9 & 946 & 431 & 341 & 289 & 41 & --- \\
\bottomrule
      \multicolumn{15}{l}{\footnotesize \textbf{Atl} = Atelectasis, \textbf{Car} = Cardiomegaly, \textbf{Con} = Consolidation, \textbf{Ede} = Edema, \textbf{Eff} = Effusion, \textbf{Emp} = Emphysema, \textbf{Fib} = Fibrosis} \\
      \multicolumn{15}{l}{\footnotesize \textbf{Her} = Hernia, \textbf{Inf} = Infiltration, \textbf{Mas} = Mass, \textbf{Nod} = Nodule, \textbf{Ple} = Pleural\_Thickening, \textbf{Pne} = Pneumonia, \textbf{Ptx} = Pneumothorax} \\
\end{tabular}

 \end{minipage}
\end{table*}

\section{Observing and Modifying Attention Supplementary}
\label{sec:attention}
\begin{figure*}[h]
\centering
\includegraphics[width=1\linewidth]{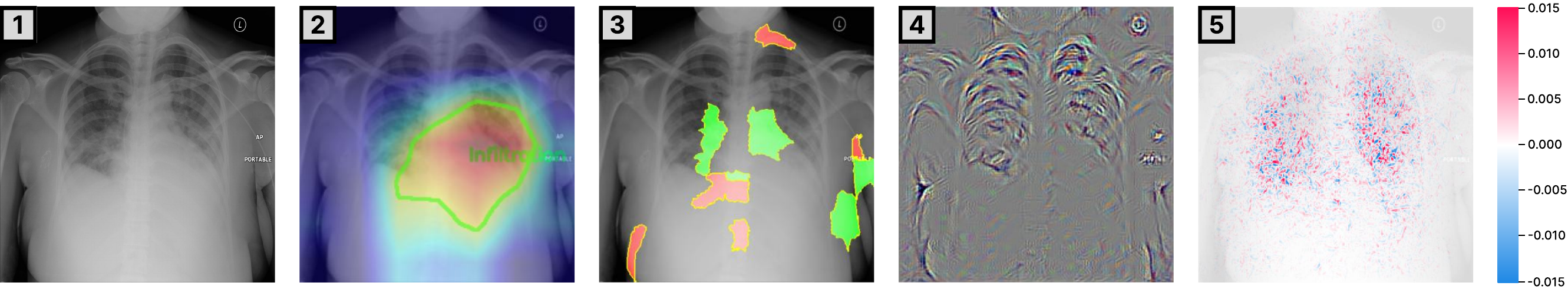}
\caption{The comparison of different visualization and interpretability methods is as follows: \raisebox{-0.07cm}{\includegraphics[height=0.35cm]{Figure/icon1.pdf}} Original Chest X-ray: Provide the baseline image for analysis. \raisebox{-0.07cm}{\includegraphics[height=0.35cm]{Figure/icon2.pdf}} Grad-CAM: Highlights the most influential regions in a coherent, high-level manner. \raisebox{-0.07cm}{\includegraphics[height=0.35cm]{Figure/icon3.pdf}} LIME (Top 10 Regions): Delivers fragmented, isolated areas that can be difficult to interpret within a broader diagnostic context. \raisebox{-0.07cm}{\includegraphics[height=0.35cm]{Figure/icon4.pdf}} Guided Backpropagation: May introduce noise, potentially obscuring clinically relevant features, particularly in multi-label settings. \raisebox{-0.07cm}{\includegraphics[height=0.35cm]{Figure/icon5.pdf}} SHAP: Often results in dispersed or sparse patterns, making interpretation more challenging.}
\label{fig:visualization}
\end{figure*}

See~\autoref{fig:visualization}.

\section{Evaluation Metrics Supplementary}
\label{sec:Metrics}

\begin{equation}
\label{equ:precision}
  precision = \frac{TP}{TP + FP}
\end{equation}
where True Positive (\textit{TP}) is the number of positive instances correctly predicted as positive; False Positive (\textit{FP}) is the number of negative instances incorrectly predicted as positive.

\begin{equation}
\label{equ:recall}
  recall = \frac{TP}{TP + FN}
\end{equation}
where False Negative (\textit{FN}) is the number of positive instances incorrectly predicted as negative.

\begin{equation}
\label{equ:F1}
  \textit{F1 score} = 2 \times \frac{precision \times recall}{precision + recall}
\end{equation}

\begin{equation}
\label{equ:AUC}
AUC = \int_{0}^{1} TPR \, d(FPR)
\end{equation}
where True Positive Rate (\textit{TPR}) is equivalent to Recall; False Positive Rate (\textit{FPR}) is:
\begin{equation}
FPR = \frac{FP}{FP + TN}
\end{equation}
where True Negative (\textit{TN}) is the number of negative instances correctly predicted as negative.

\begin{equation}
\label{equ:accuracy}
  \textit{accuracy} = \frac{TP + TN}{TP + TN + FP + FN}
\end{equation}

\begin{equation}
\label{equ:specificity}
  \textit{specificity} = \frac{TN}{FP + TN}
\end{equation}

\begin{equation}
\label{equ:map}
  \textit{mAP} = \frac{1}{N} \sum_{i=1}^{N} \textit{AP}_i
\end{equation}
where \(N\) is the total number of classes; \(\textit{AP}_i\) is the \textit{average precision} for the \(i\)-th class:
\begin{equation}
\textit{AP}_i = \frac{1}{|R_i|} \sum_{r \in R_i} P(r) \cdot \Delta r
\end{equation}
where \(R_i\) is the set of retrieved results for the \(i\)-th class, ordered by their confidence scores; \(P(r)\) is the \textit{precision} at the \(r\)-th rank; \(\Delta r\) is the change in \textit{recall} from the \((r-1)\)-th to the \(r\)-th retrieved result.

See~\autoref{tab:metrics}.

\begin{table*}[h]
    \caption{Evaluation Metrics and Justifications.}
  \label{tab:metrics}
  \begin{tabular}{lcl}
    \toprule
    \textbf{Evaluation Metrics} & \textbf{Selected} & \textbf{Justification} \\
    \midrule
   Precision~\eqref{equ:precision}                 & \Checkmark       & Focus on poor performance in underrepresented labels. \\ 
Recall/Sensitivity~\eqref{equ:recall}         & \Checkmark       & Focus on poor performance in underrepresented labels. \\ 
F1 Score~\eqref{equ:F1}                  & \Checkmark       & Balanced measure between precision and recall.  \\ 
AUC~\eqref{equ:AUC}                       & \Checkmark       & Measurement on model's discriminatory ability between classes. \\ 
Accuracy~\eqref{equ:accuracy}                  & \XSolidBrush           & Score inflation tendency due to majority labels. \\ 
Specificity~\eqref{equ:specificity}               & \XSolidBrush           & Score inflation tendency due to majority labels. \\ 
Mean Average Precision (mAP)~\eqref{equ:map}    & \XSolidBrush           & Not tailed to one specific label. \\ 
    \bottomrule
  \end{tabular}
\end{table*}

\section{User Study Supplementary}
\label{sec:User Study}
\begin{figure*}[h]
  \centering
\includegraphics[width=0.4\textwidth]{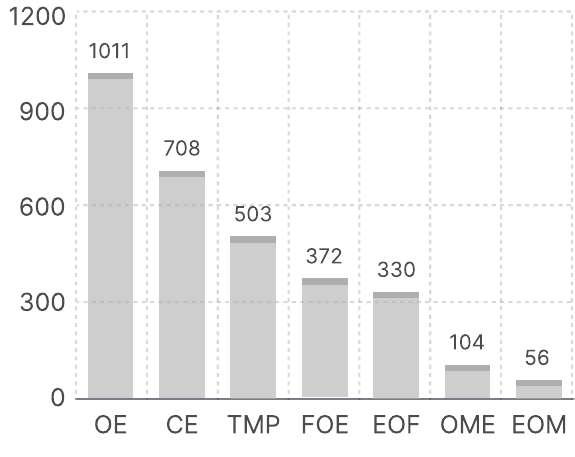}
\caption{The label distribution of \textbf{EarEndo}.}
  \label{fig:ear}
\end{figure*}

See~\autoref{fig:ear} and~\autoref{tab:earMatrix}.

\begin{table*}[h]
\begin{minipage}{\textwidth}
  \begin{center}
  \caption{The co-occurrence matrix of \textbf{EarEndo}.}
\label{tab:earMatrix}
  \begin{tabular}{cccccccc}
  \toprule
  & \textbf{OME} & \textbf{EOF} & \textbf{OE} & \textbf{EOM} & \textbf{FOE} & \textbf{CE} & \textbf{TMP} \\
  \midrule
  \textbf{OME} & --- & 0 & 10 & 0 & 0 & 0 & 0 \\
  \textbf{EOF} & 0 & --- & 41 & 0 & 0 & 26 & 8 \\
  \textbf{OE} & 10 & 41 & --- & 5 & 9 & 28 & 8 \\
  \textbf{EOM} & 0 & 0 & 5 & --- & 0 & 6 & 0 \\
  \textbf{FOE} & 0 & 0 & 9 & 0 & --- & 75 & 0 \\
  \textbf{CE} & 0 & 26 & 28 & 6 & 75 & --- & 5 \\
  \textbf{TMP} & 0 & 8 & 8 & 0 & 0 & 5 & --- \\
  \bottomrule
      \multicolumn{8}{l}{\footnotesize \textbf{OME} = Otitis Media with Effusion} \\
      
      \multicolumn{8}{l}{\footnotesize \textbf{EOF} = External Auditory Canal Foreign Body} \\

      \multicolumn{8}{l}{\footnotesize \textbf{OE} = Otitis Externa} \\

      \multicolumn{8}{l}{\footnotesize \textbf{EOM} = External Auditory Canal Tumor} \\

      \multicolumn{8}{l}{\footnotesize \textbf{FOE} = Fungal Otitis Externa} \\

      \multicolumn{8}{l}{\footnotesize \textbf{CE} = Cerumen Impaction} \\

      \multicolumn{8}{l}{\footnotesize \textbf{TMP} = Tympanic Membrane Perforation} \\
      

  \end{tabular}
  \end{center}
   \end{minipage}
\end{table*}


\end{document}